\documentclass[11pt]{article}

\usepackage{geometry}
\usepackage{amsbsy,amsmath,amsthm,amssymb,graphicx}
\usepackage{natbib}
\usepackage{enumitem}
\usepackage{pstricks}
\usepackage{tcolorbox}
\usepackage{times}

\usepackage{graphicx}
\graphicspath{ {plots/} }

\usepackage{caption}
\usepackage{subcaption}

\usepackage{amssymb}
\usepackage{epstopdf}
\usepackage[parfill]{parskip}
\usepackage{times}
\usepackage{enumerate}
\usepackage{nameref,hyperref,}
\usepackage[capitalise]{cleveref}
\usepackage[T1]{fontenc}
\usepackage[utf8]{inputenc}
\usepackage{mathtools}

\usepackage[normalem]{ulem}
\usepackage{natbib}
\usepackage{bm}
\usepackage{multirow}
\usepackage{setspace}  
\usepackage{tabulary}
\newcolumntype{K}[1]{>{\centering\arraybackslash}p{#1}}

\usepackage{times}

\newtheorem{remark}{Remark} 

\def\myfootnote{\footnote{Corresponding author. \newline
                          \hspace*{5mm} E-mail addresses:
                          z.jin@ufl.edu (Z. Jin),
                          jcsosam@unal.edu.co (J. Sosa),
                          betancourt-brenda@norc.org (B. Betancourt).}}

\title{\bf A robust Bayesian latent position approach for community detection in networks with continuous attributes}

\author{Zhumengmeng Jin$^{1,}$\myfootnote{} ,
        Juan Sosa$^2$ ,
        Shangchen Song$^3$ ,
        Brenda Betancourt$^4$\\[2ex]
    \small $^1$ University of Florida, Department of Statistics, Gainesville, FL, USA \\%
    \small $^2$ Universidad Nacional de Colombia, Bogot{\'a} D.C., Colombia \\%
    \small $^3$ University of Florida, Department of Biostatistics, Gainesville, FL, USA \\
    \small $^4$ NORC at the University of Chicago, Bethesda, MD, USA \\
    }

\date{}
\begin{document}

\maketitle

\begin{abstract}
The increasing prevalence of multiplex networks has spurred a critical need to take into account potential dependencies across different layers, especially when the goal is community detection,
which is a fundamental learning task in network analysis.
We propose
a full Bayesian mixture model for community detection in both single-layer and multi-layer networks.
A key feature of our model is the joint modeling of the nodal attributes that often come with the network data as a spatial process over the latent space.
In addition, our model for multi-layer networks allows layers to have different strengths of dependency in the unique latent position structure and assumes that the probability of a relation between two actors (in a layer) depends on the distances between their latent positions (multiplied by a layer-specific factor) and the difference between their nodal attributes.
Under our prior specifications, the actors' positions in the latent space arise from a finite mixture of Gaussian distributions, each corresponding to a cluster.
Simulated examples show that our model outperforms existing benchmark models and exhibits significantly greater robustness when handling datasets with missing values.
The model is also applied to a real-world three-layer network of employees in a law firm.
\end{abstract}


\section{Introduction}\label{sec:intro}

Network data conveniently describes the relationships between actors in complex systems \cite{kolacyzk:2009,newman:2018} and is ubiquitous in many statistical applications, including
finance \cite{Brenda:finance:2019,Gleditsch:2002},
social science \cite{Doreian:2004},
criminology \cite{McGloin:2005,Campedelli:2019},
biology \cite{Airoldi:2007},
epidemiology \cite{Volz:2009},
and computer science \cite{wang:2003}, among others.
Understanding the relationships between actors can aid domain experts.
For instance, in epidemiology, people in a certain area can be portrayed in a contact network that can be studied to detect infectious disease outbreaks \cite{Volz:2009}.
In criminology, communications between terrorists form a terrorist network, helping intelligence agencies to better counter terrorism \cite{Campedelli:2019}.

Many models have been developed for the inference of networks over the past decades (e.g., \cite{erdos:1959,ERGM:1986}),
among which the broad class of latent space models is one of the most widely used (see, e.g., \cite{sosa:2021} for an exhaustive review).
Suppose the network under study has $N$ actors, then under latent space models, there are $N$ independent and identically distributed (i.i.d.) latent variables $z_1,\dots,z_N$, one for each actor.
Under a mild exchangeability assumption in \cite{hoff:2007}, results in \cite{aldous:1985,hoover:1982} show that edge variables $y_{i,j}$, where $i,j \in \{1, \dots, N\}$, depend on latent variables through a symmetric function $\gamma(z_i, z_j)$ that is meant to capture any pattern in the network beyond any known covariate information.

Many well-known models fall into the category of latent space models, which can be distinguished between two cases depending on whether latent variables are discrete or continuous \citep{matias:2014}.
For instance, stochastic block models
\citep{nowi:2001,wang:1987}
-- hereafter SBM --
are special cases of latent space models with discrete latent variables $z_i \in \{1,2,\dots, K\}$.
When latent variables are assumed to be continuous,
another approach using latent variables is the class of latent position models (LPM) \cite{hoff:2002}
which our model in the paper is built upon.
In its basic formulation, LPMs model the edge variables $y_{i,j}$ as conditionally independent given the distance between latent variables $\gamma(z_i, z_j) = - \|z_i - z_j\|$, which naturally accounts for transitivity effects through the latent space (typically a Euclidean $K$-dimensional space for a predetermined $K$) where $z_i$ lives.
Later on, \citet{handcock:2007} proposed an extension on
LPM,
namely the latent position cluster model (LPCM),
by imposing a Gaussian mixture prior on the latent positions to perform clustering tasks.
\citet{kriv:2009} further extended
LPCM by adding the random sender and receiver effects
\cite{hoff:2005}.
Other formulations of $\gamma(\cdot, \cdot)$ can be found in \cite{schw:2003,hoff:2005,hoff:2009,athr:2017,minhas:2019}, among others.

Besides edge information of a network, extra information like node and edge attributes and different types of edges are often available, and should ideally be leveraged for inference.
Typical ways to incorporate attributes in a network model include: (1) modeling the network as a function of the attributes (see, e.g., \cite{hoff:2002,hoff:2005});
(2) modeling the attributes as a function of the network \citep{guha:2021};
(3) jointly modeling the network and attributes
\citep{link:2007,kim:2012,fosd:2015,Ciminelli}.
We consider taking the joint modeling approach similar to the social network spatial model (SNSM) proposed in \citet{Ciminelli}.
Denote the continuous nodal attribute for actor $i$ as $x_i$,
SNSM assumes that edges $y_{i,j}$ are conditionally independent given $\|z_i - z_j\|$ and $\|x_{i} - x_j\|$, and models nodal attributes as a spatial process over the latent space.
Note that joint modeling does not require the network or the attributes to be fully observed as the first two approaches do, hence one could predict missing network and attribute data (if there is any).
In addition, it improves model fitting by capturing the dependence structure between latent variables and the attributes (when such dependency exists), as we will see in \cref{sec:simulation}.

We propose a full hierarchical Bayesian model that builds on
SNSM.
But instead of using a Gaussian distribution as the prior for latent, we impose a Gaussian mixture prior so that our model could capture the group structure in the network.
Detecting communities or clusters among actors in the network is an important task in network analysis and has spurred the development of many models and algorithms,
among which the SBM has motivated an active line of research that deals with community detection
(see, e.g., \cite{Lee:2019} for a review).
However, SBM suffers from poor model fitting when many actors fall between clusters \citep{hoff:2002} as we will see in the simulation studies.
We will compare our model with an SBM that incorporates covariates as fixed effects (i.e., model the edge variables as a function of latent classes and covariates \citep{leger:2016}), and we call this model a covariate-assisted stochastic block model (CSBM).
We will show that our model presents improved model fitting while producing similar clustering results as CSBM.

We also propose an extension of our model to multi-layer network settings.
Multi-layer networks can generally be categorized into two cases: cross-sectional networks that have different types of connections
(e.g., social networks of friendship, coworker-ship, etc.) and time-varying networks where the same type of connections are measured over time (e.g., a trade network that changes over time).
We consider a type of cross-sectional multi-layer network
where each layer has a common set of actors.
Substantial work has been done on latent space models for cross-sectional
multi-layer networks that take a Bayesian approach (see, e.g.,
\cite{gollini:2016,town:2017,dangelo:2019,dangelo:2023,sosa:2022,dura:2018,wang:2019,zhu:2020}).
In extending our model to the multiple networks setting, we adopt the approach in \cite{sosa:2022}
in a parsimonious way,
where latent positions are assumed to be the same for all layers, but the strength of borrowing such latent structure information is allowed to be different across different layers.
Note that, the original model in \cite{sosa:2022} assumed different latent positions for different layers and had an additional hierarchy on the hyperparameters.
\citet{dangelo:2023} adopted a similar approach where the authors also assume the same latent positions across all layers and their model is capable of clustering assignments by using a Dirichlet process mixture on latent positions instead of a finite Gaussian mixture prior. However, their model does not take into account attribute information.
We propose a model for attributed network data sets that jointly models the network and attributes and performs clustering tasks.
The focus of the work is to see how joint modeling improves network estimation and clustering accuracy.

The remainder of the paper is organized as follows.
\cref{sec:model} contains general background on the spatial process and introduces the proposed model (in both single- and multi-layer network settings)
which we call the latent position joint mixture model (LPJMM) in the rest of the paper.
In addition,
prior specification, identifiable problem, and inference will also be discussed in this section.
Several simulation studies are conducted in \cref{sec:simulation}, where LPJMM is compared with
LPCM,
SNSM and
CSBM in single-layer settings.
We will see that LPJMM outperforms these benchmark models under different scenarios, especially when networks have missing edges.
The multi-layer model is also evaluated under a two-layer network scenario.
\cref{sec:illustration} illustrates how to use the proposed model to provide insights from a given network and the model is applied to a real-world multi-layer network data set.
Finally, we conclude with some discussion in \cref{sec:discussion}.

\section{Models}\label{sec:model}

We first review the LPM introduced in \cite{hoff:2002}, and then build upon it with a spatial process as in \cite{Ciminelli} to allow for joint modeling of the network and the nodal attributes, and with a finite Gaussian mixture distribution for latent positions to allow for clustering.

Consider a binary single-layer network with $N$ actors. Denote its adjacency matrix as $\mathbf{Y} = (y_{i,j}) \in \{0,1\}^{N\times N}$, where $y_{i,j} = 1$ if actors $i$ and $j$ are connected, and $y_{i,j} = 0$ if they are not connected.
Suppose the network data comes with a one-dimensional nodal attribute $x_i$ for each actor, and denote the covariate as $\mathbf{x} = (x_i) \in \mathbb{R}^N$.
The LPM assumes that each actor $i$ has an observed latent position $z_i$ in a
$K$-dimensional Euclidean latent space, the so-called latent space, for some $K \in \mathbb{N}$.
Let $\mathbf{z} = (z_i) \in \mathbb{R}^{N\times K}$,
then LPM models $y_{i,j}$ as conditionally independent given distances between nodal attributes as well as distances between latent positions via logistic regression.
But instead of the logistic link, we use the probit link in our model.
The analysis of probit regression models can often be facilitated by a Gibbs sampler constructed using the data augmentation approach that introduces latent variables with truncated normal distributions
\cite{albert:chib:1993}.
(See also \cite{sosa:2022} for a discussion on the choice of link functions.)
Specifically, for $i,j \in \{1, \dots, N\}$ and $i \neq j$,
\begin{equation}\label{model-level-1}
y_{i,j} \mid \mathbf{z}, \mathbf{x}, a, b, \theta
\stackrel{\text{ind}}{\sim}
\mathrm{Ber} \big( \Phi(a + b |x_i - x_j| - \theta \|z_i - z_j\|) \big) \,,
\end{equation}
where $a,b \in \mathbb{R}$ and $\theta \in \mathbb{R}^+$,
$\mathrm{Ber}(p)$ is a Bernoulli distribution that takes value 1 with some probability $p$, $\|\cdot\|$ is the Euclidean norm on $\mathbb{R}^K$ and
$\Phi(\cdot)$ is the cumulative distribution function of the standard normal distribution.
Note that we impose a factor $\theta$ for the distance between latent positions, which is different from \cite{hoff:2002} and \cite{kriv:2009}.
Although $\theta$ is unidentifiable in single-layer networks,
it plays a non-trivial role in multi-layer network settings (introduced in \cref{subsec:multi}).
We defer a detailed discussion of $\theta$ to \cref{subsec:identify}.

To allow for joint modeling of the network and nodal attributes, we model the nodal attributes as a spatial process over the latent space $\mathbb{R}^K$.
Hence, nodal attributes are treated as random variables indexed by their latent positions, and the distance between these random variables is found by the distance between their corresponding positions.
As in \cite{Ciminelli}, we specify the spatial process as a Gaussian process that is stationary with mean $\beta$ and isotropic
(see \cite{banerjee:2015} for definitions).
In this case, the process is completely defined by its covariance function
$\gamma(d)$, where $d$ is the distance between two random variables in the Gaussian process.
In particular, we specify
$\gamma(d)$ with an exponential kernel, that is,
\begin{equation*}
\gamma(d)=
        \begin{cases}
          \tau^2 + \sigma^2,              &\text{if $d = 0$;}\\
          \sigma^2 \exp(-\phi d),         &\text{if $d > 0$,}
        \end{cases}
\end{equation*}
where $\tau \geq 0$, $\sigma > 0$ and $\phi > 0$.
It is well-known that such a covariance structure is
\textit{valid}, i.e., the covariance matrix for any finite collection of random variables in the process is positive definite \citep{banerjee:2015}.
Let $M_{\mathbf{z}} = (m_{ij}) \in \mathbb{R}^{N \times N}$ where $m_{ij} = \exp(-\phi \|z_i - z_j\|)$ and denote $I_N$ as the $N$-dimensional identity matrix,
then the Gaussian process of the nodal attributes is constructed as follows,
\begin{equation}\label{model-level-2}
\mathbf{x} \mid \mathbf{z}, \beta, \sigma, \tau, \phi
\sim
\mathrm{N}_N (\beta\pmb{1}_N, \, \sigma^2 M (\mathbf{z},\phi) + \tau^2 I_N),
\end{equation}
where $\mathrm{N}_d$ is a $d$-dimensional multivariate normal distribution for some dimension $d \in \{2, 3, \dots\}$,
and $\pmb{1}_N$ is an $N$-dimensional vector with all 1s.

We then impose a Gaussian mixture distribution on latent positions, which allows us to cluster actors into different groups.
Suppose there are $H < \infty$ predetermined number of components in the Gaussian mixture distribution, then
\begin{equation}\label{model-level-3}
z_i \mid \pmb{\omega}, \pmb{\mu}, \pmb{\kappa}
\stackrel{\text{i.i.d.}}{\sim}
\sum_{h=1}^H \omega_h  \mathrm{N}_K (\mu_h, \, \kappa_h^2 I_K)\, ,
\end{equation}
where $\pmb{\omega} = \{\omega_1, \dots, \omega_H\}$, $\pmb{\mu} = \{\mu_1, \dots, \mu_H\}$, $\pmb{\kappa} = \{\kappa_1, \dots, \kappa_H\}$.
Note that $\mu_h$ is a $K$-dimensional mean vector where $h \in \{1, \dots, H\}$,
and $\omega_h$ is the probability that an actor belongs to the $h$-th group such that $\omega_h \in (0,1)$ and $\sum_{h=1}^H \omega_h = 1$.

In single-layer network settings, the model is given by
\cref{model-level-1,model-level-2,model-level-3}.
Under our model, nodal attributes of two actors whose latent positions are close are more likely to be similar according to the exponential covariance structure.
If $b < 0$ ($b > 0$), actors with similar attributes are more (less) likely to be connected.
When $b = 0$, nodal attributes do not affect the distribution of the network directly (but it still has an indirect impact on the network through latent positions by \cref{model-level-2}).

\subsection{An extension to multi-layer networks.}\label{subsec:multi}

\begin{figure}
\centering
\includegraphics[width=.6\linewidth]{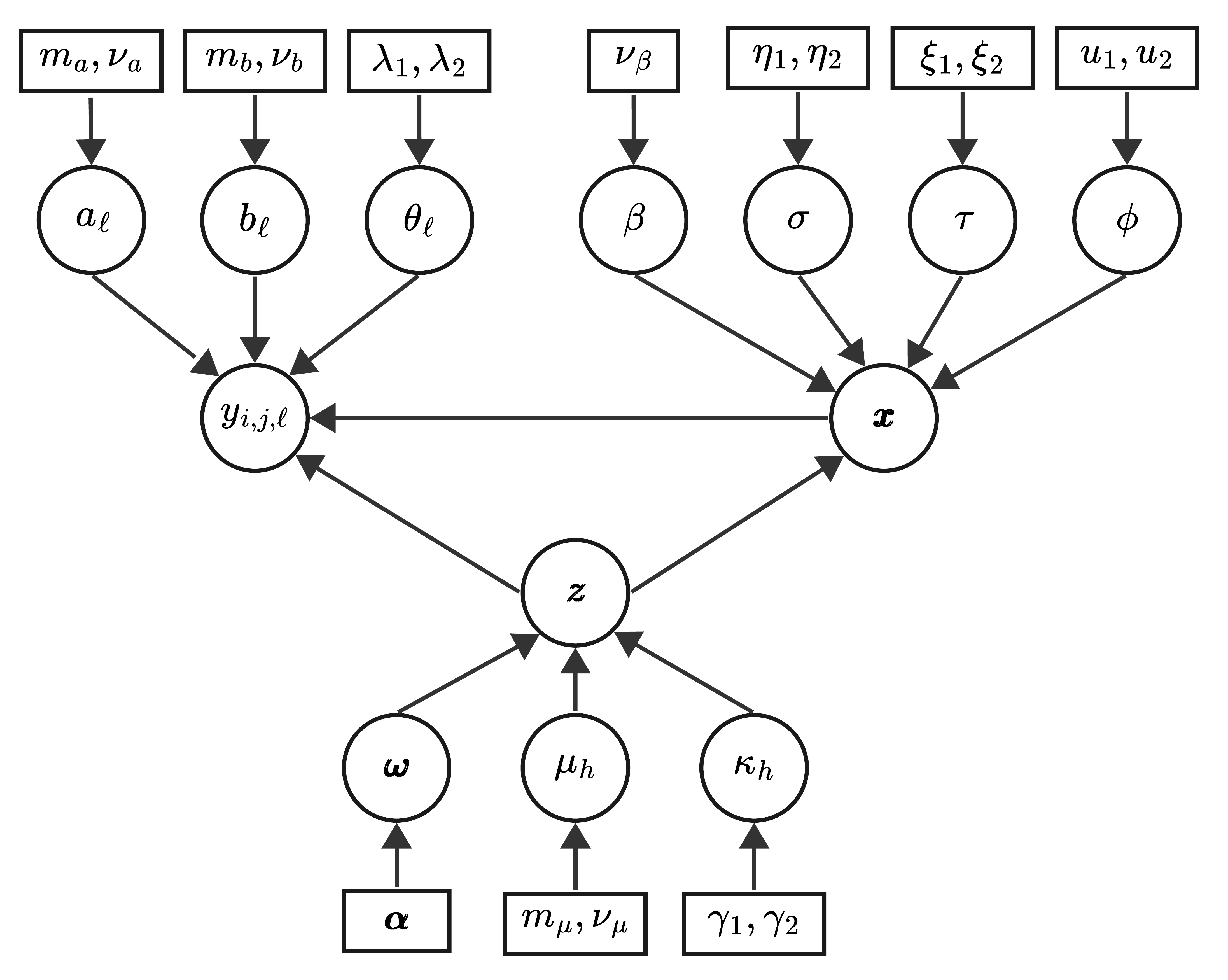}
\caption{DAG representation of the LPJMM in multi-layer settings.}
\label{fig:DAG}
\end{figure}

Our model can also be extended to multi-layer network settings in the following way.
Suppose we have $L$ layers $\mathbf{Y}_1, \dots, \mathbf{Y}_L$ in the network, where all layers are defined over the same set of actors.
We assume the same latent positions $\mathbf{z}$ for all layers but allow the strength of borrowing such latent structure information to be different by imposing layer-specific factors $\theta_\ell$ for $\ell \in \{1, \dots, L\}$.
Our model in multi-layer settings is then presented as follows
\begin{align}
y_{i,j,\ell} \mid \mathbf{z}, \mathbf{x}, a_\ell, b_\ell, \theta_\ell
& \stackrel{\text{ind}}{\sim}
\mathrm{Ber} \big( \Phi(a_\ell + b_\ell |x_i - x_j| - \theta_\ell \|z_i - z_j\|) \big)  \, ,
\label{multi-layer-1}  \\
\mathbf{x} \mid \mathbf{z}, \beta, \sigma, \tau, \phi
& \sim
\mathrm{N}_N (\beta\pmb{1}_N, \, \sigma^2 M (\mathbf{z},\phi) + \tau^2 I_N) \, ,
\label{multi-layer-2} \\
z_i \mid \pmb{\omega}, \pmb{\mu}, \pmb{\kappa}
& \stackrel{\text{i.i.d.}}{\sim}
\sum_{h=1}^H \omega_h  \mathrm{N}_K (\mu_h, \, \kappa_h^2 I_K)\, ,
\label{multi-layer-3}
\end{align}
where $y_{i,j,\ell}$ is the edge variable between actors $i$ and $j$ in layer $\ell$, and
$a_\ell, b_\ell$ and $\theta_\ell$ are layer-specific parameters.
Note that \cref{multi-layer-2,multi-layer-3} are the same as \cref{model-level-2,model-level-3}.
\cref{fig:DAG} shows a directed acyclic graph (DAG) representation of the model given by \cref{multi-layer-1,multi-layer-2,multi-layer-3}.

\subsection{Prior specification}\label{subsec:prior}

We take a Bayesian approach to estimate the model parameters.
Without loss of generality,
a Bayesian version of the model given by
\cref{multi-layer-1,multi-layer-2,multi-layer-3}
is formed by placing prior distributions on the unknown parameters
$a_\ell, b_\ell, \theta_\ell$,
$\beta$,
$\sigma$,
$\tau$,
$\phi$,
$\pmb{\omega}$,
$\pmb{\mu}_h, \kappa_h$,
for $\ell = \{1, \dots, L\}$ and $h = \{1, \dots, H\}$.
In the model we consider, these parameters are assumed \textit{a priori} independent.
For parameters in the probit regression tier as specified by \cref{multi-layer-1}, their priors are specified as follows:
\begin{align*}
a_\ell  \stackrel{\text{i.i.d.}}{\sim} \mathrm{N} (m_a, \nu_a^2)\, ,
\qquad
b_\ell  \stackrel{\text{i.i.d.}}{\sim} \mathrm{N} (m_b, \nu_b^2)\, ,
\qquad
\theta_\ell \stackrel{\text{i.i.d.}}{\sim} \mathrm{Gamma} (\lambda_1, \lambda_2) \, ,
\end{align*}
for $\ell \in \{1, \dots, L\}$.
The priors for the parameters in the spatial process tier as given in \cref{multi-layer-2} are given as follows:
\begin{align*}
\beta    \sim \mathrm{N} (0, \nu_\beta^2) \, ,
\qquad
\sigma^2 \sim \mathrm{InvG} (\eta_1, \eta_2) \, ,
\qquad
\tau^2   \sim \mathrm{InvG} (\xi_1, \xi_2) \, ,
\qquad
\phi     \sim \mathrm{U} (u_1, u_2) \, .
\end{align*}
Finally, we put the following priors on the rest of the parameters:
\begin{align*}
\pmb{\omega} \sim \mathrm{Dir} (\alpha) \, ,
\qquad
\mu_h \stackrel{\text{i.i.d.}}{\sim} \mathrm{N}_K (m_\mu, \nu_\mu^2 I_K) \, ,
\qquad
\kappa^2_h \stackrel{\text{i.i.d.}}{\sim} \mathrm{InvG} (\gamma_1, \gamma_2) \, .
\end{align*}
Note that, $m_a$, $\nu_a$, $m_b$, $\nu_b$, $\lambda_1$, $\lambda_2$, $\nu_\beta$, $\eta_1$, $\eta_2$, $\xi_1$, $\xi_2$, $u_1$, $u_2$, $\alpha$, $m_\mu$, $\nu_\mu$, $\gamma_1$ and $\gamma_2$ are
user-specified hyperparameters,
and $\mathrm{Gamma}(\cdot, \cdot)$,
    $\mathrm{InvG} (\cdot, \cdot)$,
    $\mathrm{U}(\cdot, \cdot)$,
    $\mathrm{Dir}(\cdot)$ represents Gamma, Inverse-Gamma, uniform, and Dirichlet distributions respectively.

\subsection{Posterior distribution and model estimation}\label{subsec:posterior}
As is standard in Bayesian estimation of mixture models
(see, e.g., \cite{diebolt:1994}),
we define a new variable $g_i$ that serves as the missing data of group membership of actor $i$ whose distribution depends on $\pmb{\omega}$. In particular, $g_i = h$ if actor $i$ belongs to the $h$-th group.
The joint density of $(z_i, g_i)$ given $\pmb{\omega}$, $\pmb{\mu}$ and $\pmb{\kappa}$ is then given by
\[
\prod_{h=1}^H \bigg\{
\omega_h
\frac{1}{\sqrt{2\pi\kappa_h^2}}
\exp \Big(
-\frac{1}{2 \kappa_h^2} \|z_i - \mu_h\|^2
\Big) \bigg\}^{\mathrm{I}_{\{g_i = h\}}} \, ,
\]
where the indicator function $\mathrm{I}_{\{g_i = h\}} = 1$ if $g_i = h$, and $\mathrm{I}_{\{g_i = h\}} = 0$ otherwise.
Let $\mathbf{g} = (g_i)_{i=1}^N$ be the group membership for all actors
and $\mathcal{L}(\cdot)$ be the law of a random variable.
Then the posterior distribution of $\mathbf{z}$, $\mathbf{g}$ and the parameters (whose priors are specified in \cref{subsec:prior}) is given by
\begin{align*}
\Pi( & \mathbf{z}, \mathbf{g}, a_1, \dots, a_L, b_1, \dots, b_L, \theta_1, \dots, \theta_L, \beta, \tau^2, \sigma^2, \phi, \pmb{\omega},  \pmb{\mu}, \pmb{\kappa} \mid \mathbf{Y}_1, \dots, \mathbf{Y}_L, \mathbf{x}) \\
\propto &
  \bigg\{ \prod_{\ell = 1}^L \mathcal{L} (\mathbf{Y}_\ell \mid \mathbf{z}, \mathbf{x}, a_\ell, b_\ell, \theta_\ell) \bigg\}
  \mathcal{L} (\mathbf{x} \mid \mathbf{z}, \sigma, \tau, \phi)
  \mathcal{L} (\mathbf{z}, \mathbf{g} \mid \pmb{\omega}, \pmb{\mu}, \pmb{\kappa})
  \bigg\{\prod_{\ell=1}^L \mathcal{L} (a_\ell) \mathcal{L} (b_\ell) \mathcal{L} (\theta_\ell) \bigg\} \\
& \times
  \mathcal{L} (\beta)
  \mathcal{L} (\sigma^2)
  \mathcal{L} (\tau^2)
  \mathcal{L} (\phi)
  \mathcal{L} (\pmb{\omega})
  \mathcal{L} (\pmb{\mu})
  \mathcal{L} (\pmb{\kappa}) \, .
\end{align*}
Note that the dimension of the posterior distribution has dimension $NK + N + 3L + 3H + 4$
and the corresponding posterior density is presented as follows,
\begin{align*}
\pi(& \mathbf{z}, \mathbf{g}, a_1, \dots, a_L, b_1, \dots, b_L, \theta_1, \dots, \theta_L, \beta, \tau^2, \sigma^2, \phi, \pmb{\omega},  \pmb{\mu}, \pmb{\kappa} \mid \mathbf{Y}_1, \dots, \mathbf{Y}_L, \mathbf{x})\\
\propto &
\bigg\{
\prod_{\substack{i,j=1 \\ i\neq j}}^N \prod_{\ell = 1}^L
  \big[\Phi(a_\ell + b_\ell |x_i - x_j| - \theta_\ell \|z_i - z_j\|)\big]^{y_{i,j,\ell}}
  \\
  & \hspace{15mm} \times \big[1-\Phi(a_\ell + b_\ell |x_i - x_j| - \theta_\ell \|z_i - z_j\|)\big]^{1-y_{i,j,\ell}}
  \bigg\}\\
& \times
  |\sigma^2 M(\mathbf{z}, \phi) + \tau^2 I_N| ^{-\frac{1}{2}}
  \exp\Big(-\frac{1}{2}(\mathbf{x} - \beta \mathbf{1})^\intercal \big(\sigma^2 M(\mathbf{z}, \phi) + \tau^2 I_N\big)^{-1} (\mathbf{x} - \beta \mathbf{1}) \Big) \\
& \times
  \prod_{i=1}^N \prod_{h=1}^H \bigg\{
  \frac{\omega_h}{\sqrt{\kappa_h^2}}
  \exp \Big(
       -\frac{1}{2 \kappa_h^2} \|z_i - \mu_h\|^2
       \Big)
  \bigg\}^{\mathrm{I}_{\{g_i = h\}}}\\
& \times
  \exp\Big(  \frac{1}{2 \nu_a^2}\sum_{\ell=1}^L(a_\ell-m_a)^2
           + \frac{1}{2 \nu_b^2}\sum_{\ell=1}^L(b_\ell-m_b)^2
      \Big)
  \prod_{\ell=1}^L \theta_\ell^{\lambda_1-1} \exp(-\lambda_2 \theta_\ell)\\
& \times
  \exp\Big(\frac{\beta^2}{2\nu_\beta^2}\Big)
  (\sigma^2)^{-\eta_1-1}  (\tau^2)^{-\xi_1-1}
  \exp\Big(-\frac{\eta_2}{\sigma^2} - \frac{\xi_2}{\tau^2}\Big)
  \mathrm{I}_{\{\phi\in [u_1, u_2]\}} \\
& \times
  \prod_{h=1}^H \bigg\{
  \omega_h^{\alpha_h-1}  \mathrm{I}_{\{\sum_{h=1}^H\omega_h=1\}}
  \exp\Big( -\frac{1}{2\nu_\mu^2} \|\mu_h - m_\mu\|^2 \Big)
  (\kappa_h^2)^{-\gamma_1 - 1} \exp\Big(-\frac{\gamma_2}{\kappa_h^2}\Big)
  \bigg\}.
\end{align*}

\subsection{Inference and identifiability of parameters}\label{subsec:identify}

Note that the posterior distribution is highly intractable, hence we must resort to Markov chain Monte Carlo (MCMC) methods for inferences on model parameters.
A Markov chain of the parameters is generated via the program ``Just Another Gibbs Sampler'' (JAGS)
which is implemented
in \texttt{R} \citep{R:2021} using the $\mathtt{rjags}$ package \citep{rjags:2022}.

Several parameters are not identifiable in our model.
Firstly, due to factors $\theta_\ell$ and $\phi$, and the fact that latent positions are incorporated in the posterior only through their distances,
the posterior is, therefore, invariant to $\theta_\ell$s and $\phi$, and is invariant to scaling, reflection, rotation, and translation of the latent positions $\mathbf{z}$.
(Note that, \cite{hoff:2002,kriv:2009} did not have $\theta_\ell$s, hence their posterior is not invariant to the scaling of latent positions.)
Although $\theta_\ell$s are not identifiable and do not affect the model fitting, their ratios $\theta_{\ell1} / \theta_{\ell2}$ still provide valid information on layer's relative strength of borrowing information from the latent space in multi-layer settings.

Despite being unidentifiable, one can still make inferences on the latent positions and find a reasonable estimate for $\mathbf{z}$ through a post-process which we now describe.
Similar to the definition in \cite{hoff:2002}, we define the equivalence class of $\mathbf{z} \in \mathbb{R}^{N \times K}$, denoted as $[\mathbf{z}]$, to be the set of positions that are equivalent to $\mathbf{z}$ under scaling, reflection, rotation, and translation.
Given a fixed reference position $\mathbf{z}_{ref}$, a position $\mathbf{z}_{\ast}$ is found in $[\mathbf{z}]$ such that
$\mathbf{z}_{\ast} = \arg\,\min_{\mathbf{z}' \in [\mathbf{z}]} \mathrm{tr}(\mathbf{z}_{ref} - \mathbf{z}')^\intercal (\mathbf{z}_{ref} - \mathbf{z}')$, which is the so-called Procrustes transformation.
In simulation studies, $\mathbf{z}_{ref}$ is naturally chosen to be the true latent position, while in practical applications, we could use the last iteration of the Markov chain of latent positions as the reference.
The Procrustes transformation is performed for each iteration of the Markov chain of the latent positions $\{\mathbf{z}_n\}$, and an estimate for $\mathbf{z}$ is taken as the mean of the Procrustes transformations of $\{\mathbf{z}_n\}$.

As occurs in Bayesian mixture models, the label-switching problem for the group membership $\mathbf{g}$ is another source of non-identifiability.
That is, the posterior is invariant under permutations of clustering labels.
Many algorithms have been proposed to obtain a single clustering estimate based on the MCMC sample of the group membership $\{\mathbf{g}_n\}$, including
an optimization method (which we call ``MaxPEAR'' hereafter) that finds a clustering that maximizes posterior expected adjusted Rand index (ARI)
\cite{maxpear:2009},
an optimization method (``MinBinder'') that minimizes Binder's loss function
\cite{minbinder:2007},
and a greedy algorithm (``GreedyEPL'') that aims to minimize the variation of information
\cite{GreedyEPL:2018},
among others.
These approaches may generate different clustering estimates, and to get a better understanding of the model performance, all aforementioned algorithms (MaxPEAR, MinBinder, and GreedyEPL) are used to assess the model.
Estimates based on these approaches are found using the packages
$\mathtt{GreedyEPL}$ \citep{pkg:GreedyEPL}
and $\mathtt{mcclust}$ \citep{pkg:mcclust}.

\section{Simulation}\label{sec:simulation}

Four network scenarios are considered in this section to evaluate our model.
The first three simulations consider single-layer networks generated from LPJMM, but the underlying latent positions display varying degrees of clustering.
In each of the three simulations,
LPJMM is compared with three other models designed only for single-layer networks, namely
LPCM in \cite{handcock:2007},
SNSM in \cite{Ciminelli}, and
CSBM in \cite{leger:2016}.
Model assessments include how well a model could recover the group membership and the latent position configuration, and a goodness-of-fit test using summaries of networks including density, transitivity, and assortative coefficient based on the estimated group membership $\mathbf{g}$
(see \cite{kolaczyk:2020} for definitions).
Furthermore, LPJMM is also evaluated by how accurately it estimates certain parameters.
In the last simulation, we consider a two-layer network and the performance of LPJMM could be further evaluated by the estimated ratio $\theta_1/\theta_2$, which reflects differences in each layer's dependency on the latent position structure.

LPJMM and SNSM are implemented using the $\mathtt{rjags}$ package, and LPCM and CSBM are implemented using the
$\mathtt{latentnet}$ \citep{pkg:latentnet}
and $\mathtt{sbm}$   \citep{pkg:sbm}
packages respectively.
Model specifications of these models can be found in Appendix~\ref{app:other_models}.

\subsection{A single-layer network}\label{subsec-sim-1}

\begin{figure}
\centering
\includegraphics[width=.6\linewidth]{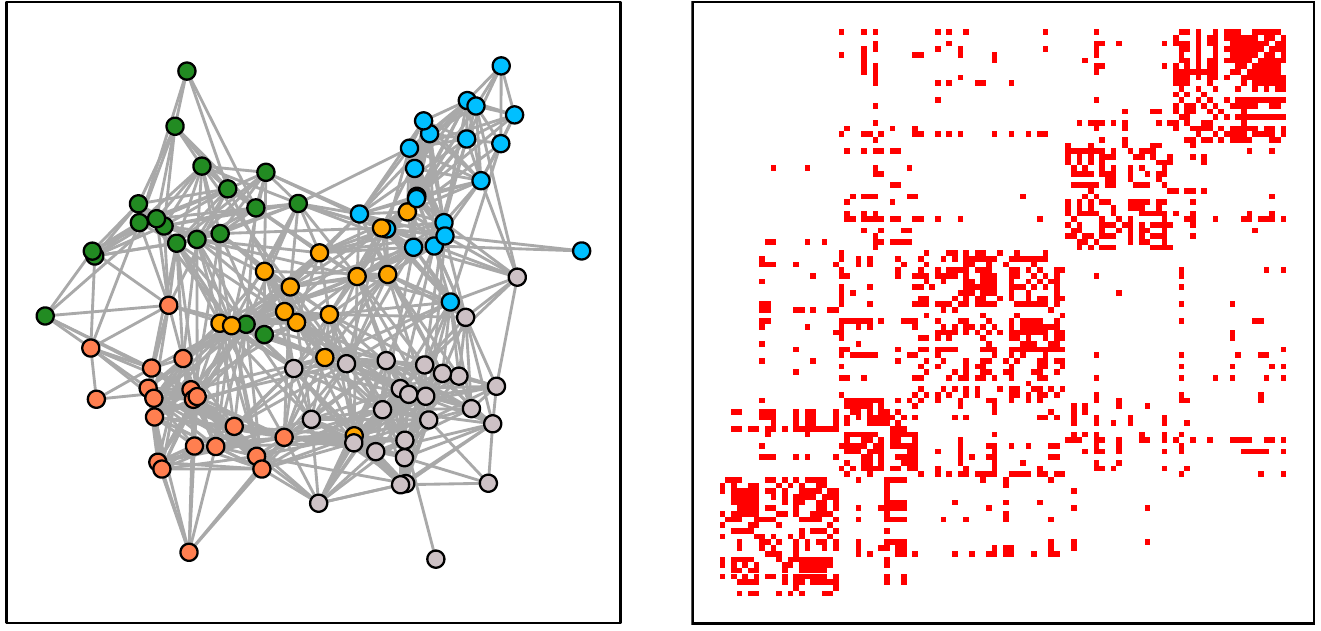}
\caption{Left: A visualization of the network based on the true latent position and color indicates group membership $\mathbf{g}$.
Right: Heatmap of the adjacency matrix (where actors are reordered according to $\mathbf{g}$).}
\label{fig:sim1-network}
\end{figure}

Consider a single-layer network (i.e., $L = 1$) with $N = 100$ actors generated as follows.
Firstly, generate latent positions $\mathbf{z}$ from a mixture of $H = 5$ multivariate normal distributions, and then generate attributes $\mathbf{x}$ jointly from a multivariate normal distribution with mean $\beta \mathbf{1}_N = \mathbf{0}$ and covariance matrix given by
$\gamma(\cdot)$
in \cref{sec:model} where $\phi = 0.5$, $\tau^2 = 0.3$, $\sigma^2 = 1$.
Finally, the network data is generated according to \cref{model-level-1} with $a = 5$, $b = -2$, and $\theta = 2.72$.
See
\cref{fig:sim1-network}
for a visualization of the simulated network.
The network with moderate density of $0.1531$ shows strong transitivity and assortative mixing with coefficients $0.5049$ and $0.5512$ respectively.
Note that the clustering pattern in the true latent space $\mathbf{z}$ is discernible with little overlap between different groups. Of course, the estimation of $\mathbf{z}$, hence $\mathbf{g}$, will be more accurate when clusters are far away from each other.
On the other hand, when members are thoroughly blended, the probabilities of connections within clusters are comparable to those across clusters, making it more difficult to estimate true group membership.
We explore these two extreme scenarios (one with a very distinctive clustering pattern and another with well-blended members) in \cref{subsec-sim-2,subsec-sim-3}.

As for the prior specifications, we set $m_a = m_b = 0$, and $\nu_a^2 = \nu_b^2 = 9$ to allow a wide range of values for $a$ and $b$.
Let $\theta \sim \mathrm{Gamma}(1,1)$ so that $\theta$ has mean 1.
An almost flat prior is imposed on $\beta$ by setting $\nu_\beta = 10^4$.
The same uniform prior $\mathrm{U}(0,1)$ as in \cite{Ciminelli} is specified for $\phi$.
We suggest the sum of the prior means of $\tau^2$ and $\sigma^2$ to be on the same scale as the sample variance of $\mathbf{x}$, and here we use
$\sigma^2 \sim \mathrm{InvG} (2, 1)$ and
$\tau^2  \sim \mathrm{InvG} (2, 1)$.
Let $\alpha = \mathbf{1}$ so that the prior on $\pmb{\omega}$ is a flat Dirichlet distribution.
Following the heuristics in \cite{sosa:2022}, we specify
$\mu_h \stackrel{\text{i.i.d.}}{\sim} \mathrm{N}_K (\mathbf{0}, 2/3 I_K)$ and
$\kappa^2_h\stackrel{\text{i.i.d.}}{\sim} \mathrm{InvG} (3, 2/3)$
so that $\mathrm{var}(z_{ij}|g_i) = 1$.

Note that the latent space dimension $K$ and the number of clusters $H$ in the model need to be prespecified along with the priors.
We take $K$ to be the true dimensions of the latent space (i.e., $K = 2$) since this facilitates model assessment by allowing visualizations of the estimated latent positions.
One could also use the Watanabe-Akaike Information Criterion (WAIC)
to select a $K$ with the smallest WAIC as in \cite{sosa:2022}.
However, WAIC and other information criteria like the Deviance Information Criterion (DIC) do not help choose the number of clusters $H$.
A comparison of the model assessment for different specified $H$ is given in Appendix~\ref{app:compare-H}.
We noticed that model performances are significantly worse when $H$ is chosen to be smaller than the truth.
However, model performances are similar among models whose $H$ is at least as large as the truth.
Therefore, we suggest choosing $H$ to be the largest number of groups that one is willing to accept, and in this example, we choose $H$ to be $5$.

We then fit LPJMM using MCMC sampling with $20\,000$ burn-in iterations and a further $10\,000$ iterations kept for posterior analysis. Before running the Markov chain, another $20\,000$ iterations for adaption was used to help the $\mathtt{rjags}$ package to choose the optimal Markov chain sampler.
The Markov chain mixes reasonably well and shows no signs of lack of convergence (see Appendix~\ref{app:traceplots} for the traceplot of the log-likelihood chain).

\begin{figure}
\centering
\includegraphics[width=.95\linewidth]{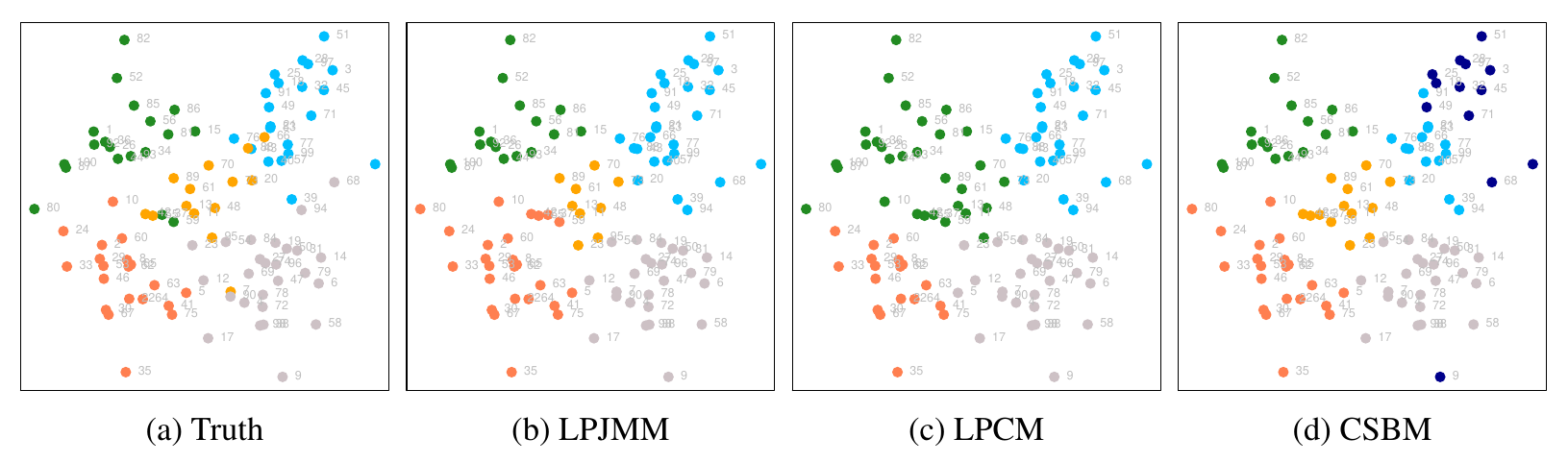}
\caption{(a): Color indicates the true group membership $\mathbf{g}$.
(b)-(d): Color indicates the estimated group memberships $\hat{\mathbf{g}}$ of LPJMM, LPCM and CSBM respectively. Positions of the points in all plots are true latent positions $\mathbf{z}$.}
\label{fig:sim1-z}
\end{figure}

To evaluate a model's ability to recover the group membership, we first find estimates of clustering using the optimization algorithms
(i.e., MaxPEAR, MinBinder, and GreedyEPL)
mentioned in \cref{subsec:identify} and find ARI for each clustering estimate.
Note that clusters are not defined in SNSM, therefore we only compare the ARI between LPJMM, LPCM, and CSBM.
Since the $\mathtt{sbm}$ package takes a non-Bayesian approach that uses a Variational-EM algorithm to find a point estimator for the group membership $\mathbf{g}$, optimization methods like MaxPEAR are not necessary to analyze results from CSBM.
Results in \cref{table:rand-index} suggest that these three models have a similar ability to recover group membership $\mathbf{g}$, with ARI of LPJMM using the MaxPEAR and MinBinder algorithms being higher than ARI ($0.707$) under the CSBM model.
Although the highest ARI is given by LPCM using the MinBinder method, the estimated number of groups is 10 which significantly exceeds the true number of groups. Therefore, the overall best estimation in terms of both ARI and $\hat{\mathbf{g}}$ is given by LPJMM.
A visualization of the estimated clusters based on the true latent positions is given in \cref{fig:sim1-z}.
Also, notice that the MinBinder algorithm tends to overestimate the number of clusters in the network in all models.

{\renewcommand{\arraystretch}{1.1}
\begin{table}
\centering
{\begin{tabular}{K{3cm}K{2cm}K{2cm}K{2cm}}
\hline
             &  LPJMM  &  LPCM       &  CSBM     \\
\hline
 MaxPEAR          & 0.737 (5)   &  0.707 (4)  &  --       \\
 MinBinder        & 0.712 (11)  &  0.748 (10) &  --       \\
 GreedyEPL        & 0.664 (4)   &  0.688 (4)  &  --       \\
 Variational-EM   & --          & --          & 0.707 (6) \\
\hline
\end{tabular}}
\label{table:rand-index}
\vspace{0.3cm}
\caption{ARI and numbers of estimated groups (in parentheses).}
\end{table}}

To further compare the ability to recover latent position configuration between LPJMM and LPCM, we find an estimate of the latent positions $\hat{\mathbf{z}}$ using the method of Procrustes transformation given in \cref{subsec:identify}.
Plots of $\hat{\mathbf{z}}$s of LPJMM and LPCM can be found in Appendix~\ref{app:compare-two-models},
which suggest similar estimated configurations of $\mathbf{z}$.

\begin{figure}
\centering
\includegraphics[width=0.9\textwidth]{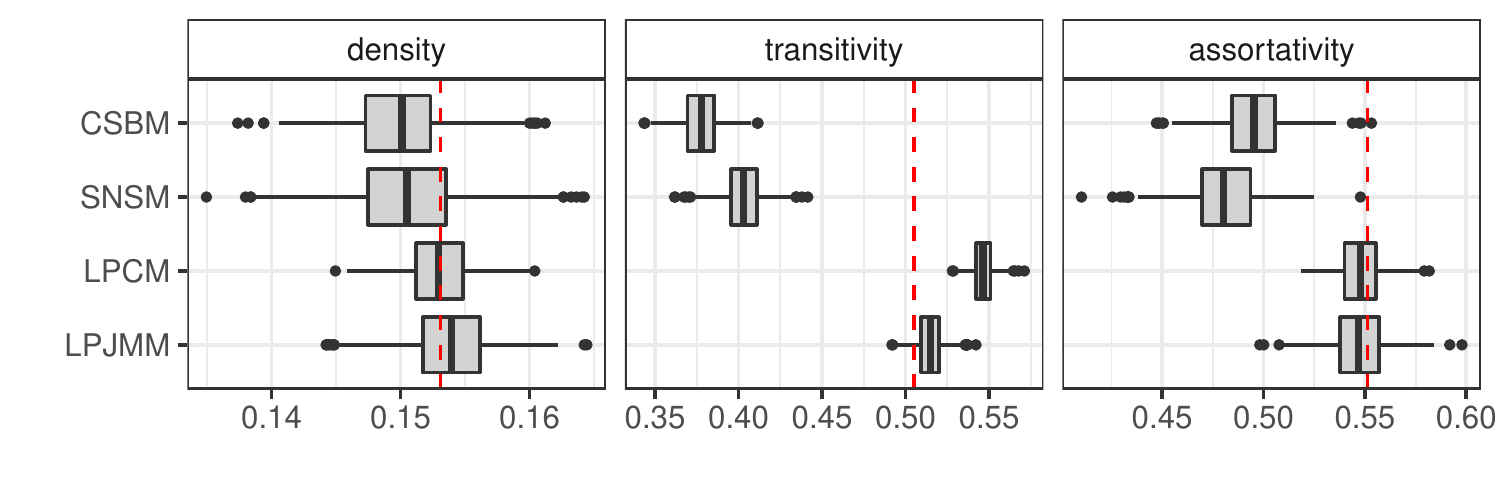}
    \caption{Boxplots of summary statistics for each model. Red dotted lines indicate the true values for network characteristics respectively.}
    \label{fig:boxplot_GoF}
\end{figure}

Following \cite{sosa:2022}, we assess if models have a good fit in the sense of good reproduction of a variety of summary statistics.
These statistics are calculated based on the networks generated from the model estimations.
For LPJMM and SNSM, networks are simulated based on the parameters at every $10$-th iteration of the generated Markov chain.
For LPCM and CSBM, $1000$ networks are simulated using their respective packages.
Then for each model, we calculate the density, transitivity, and assortative coefficient (if applicable) calculated according to the true group membership for each simulated network.
Boxplots and the averages of these summary statistics are given in
\cref{fig:boxplot_GoF,table:summary-statistics} respectively.
In general, the summary statistics results are not too disparate for all four models.
However, LPJMM better captures these structural features of the network data, while LPCM tends to overestimate transitivity in the network, and both SNSM and CSBM tend to underestimate transitivity and assortativity in the network.

{\renewcommand{\arraystretch}{1.1}
\begin{table}
\centering
{\begin{tabular}{lccccc}
 \hline
          & true value  & LPJMM    & LPCM   & SNSM    & CSBM      \\
\hline
density       & 0.1531 & 0.1539   & 0.1530 & 0.1504  & 0.1499 \\
 transitivity  & 0.5049 & 0.5144   & 0.5467 & 0.4027  & 0.3776 \\
 assortativity & 0.5512 & 0.5468   & 0.5475 & 0.4811  & 0.4954 \\
\hline
\end{tabular}}
\label{table:summary-statistics}
\vspace{0.3cm}
\caption{Means of the summary statistics of the simulated networks for each model.}
\end{table}}

{\remark The reader may wonder (as did a referee) whether there would be collinearity issues in \cref{model-level-1} due to the relationship between $x_i$ and $z_i$ given in \cref{model-level-2}. Note that in the model specification, the nodal attributes vector is modeled as a Gaussian process, where the covariance between $x_i$ and $x_j$ is given by $\sigma^2 \exp(-\phi \|z_i - z_j\|)$. This reflects our assumption that $x_i$ and $x_j$ is likely to be more similar if they are closer in the latent space (i.e., a smaller value of $\|z_i - z_j\|$ will result in a larger value of covariance between $x_i$ and $x_j$).
However, we let $\|z_i - z_j\|$ affects $|x_i - x_j|$ via the covariance matrix through a non-linear exponential function so that the model can capture a great degree of variability in the relationship between $|x_i - x_j|$ and $\|z_i - z_j\|$, thus ensuring a probabilistic relationship between them.
Therefore, we do not anticipate a linear relationship between $\|z_i - z_j\|$ and $|x_i - x_j|$. The correlation between $\|z_i - z_j\|$ and $|x_i - x_j|$ in this simulation study is found to be $0.0555$. Therefore collinearity is not a issue in the simulation study.}

Two more extreme scenarios are also considered. In the first scenario, assume all actors take the same position in the latent space (i.e., $\|z_i - z_j\| = 0$ for all $i,j$). In this case, the correlation between $\|z_i - z_j\|$ and $|x_i - x_j|$ is $0.1222$.
In another scenario, the correlation coefficient is $0.2295$ if distances between the five clusters in the latent space are even further from each other (as is the case in \cref{subsec-sim-2} below). In either case, the correlation between $\|z_i - z_j\|$ and $|x_i - x_j|$ is weak. We thus believe that collinearity should not be a problem for the model.

\subsection{A network with missing edges}
\label{subsec-sim-2}

\begin{figure}
\centering
\includegraphics[width=0.6\textwidth]{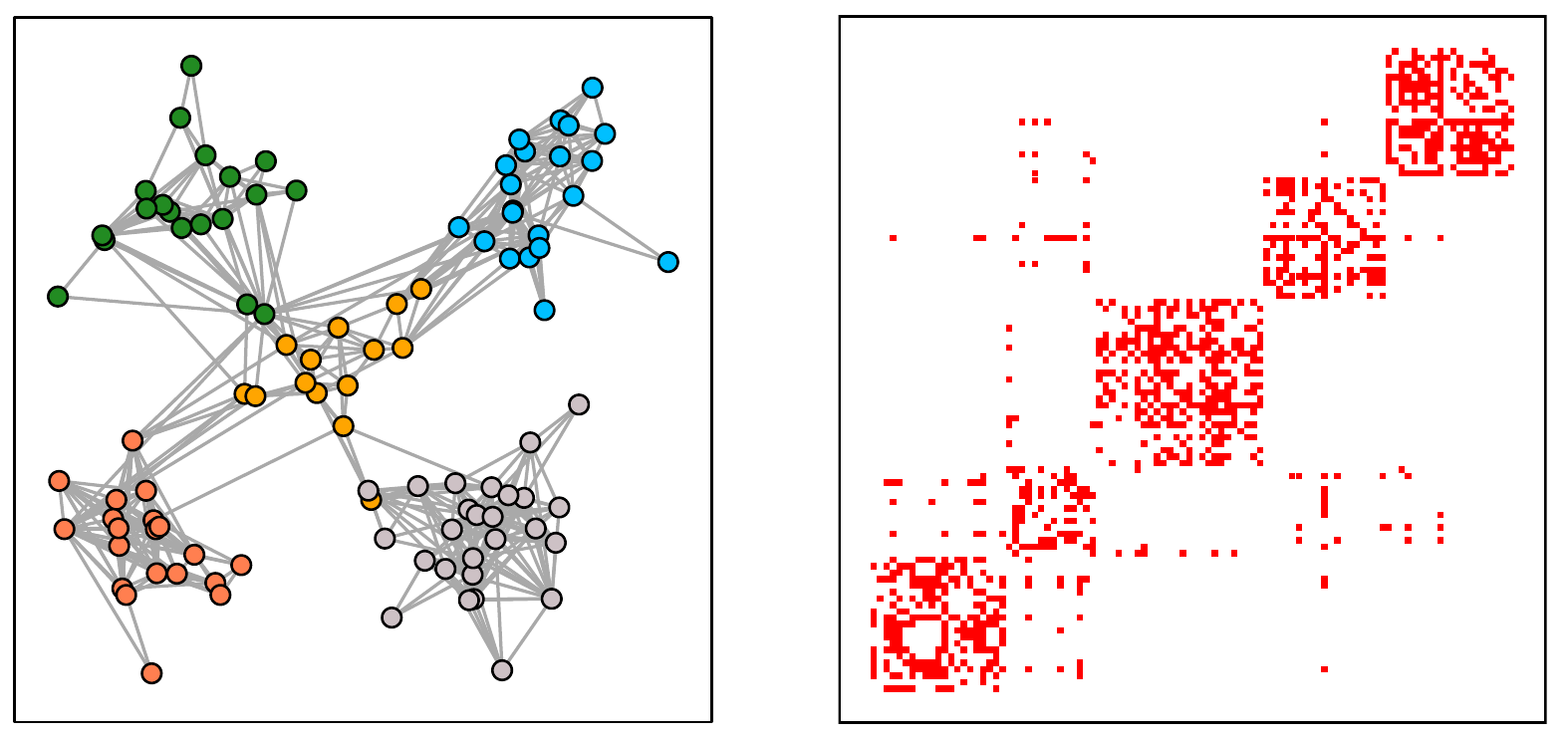}
\caption{Network visualization and heatmap of the adjacency matrix.}
\label{fig:network-far}
\end{figure}

In this simulation scenario, a network with $N = 100$, $H = 5$ is generated, which shows highly discernible clustering patterns (as shown in \cref{fig:network-far}). The model parameters used to generate the network are given by $a = 1$, $b = 3.5$, $\theta = 1.25$, $\phi = 0.3$, $\tau^2 = 0.2$, $\sigma^2 = 1$ and $\beta = 0$.
Model specifications are determined similarly as in \cref{subsec-sim-1}.
We will show that LPJMM clearly outperforms other models when the network exhibits missing edges.

\cref{table:rand-index-far} shows that,
when edge information is fully observed, all models can accurately estimate the group membership (not considering SNSM), with LPJMM slightly outperforming the rest of the two models.
Results from \cref{table:summary-statistics-far} are consistent with what we have seen in \cref{subsec-sim-1}, where LPJMM and LPCM can reproduce the summary statistics quite close to the true values, while SNSM and CSBM tend to underestimate transitivity and assortativity.

{\renewcommand{\arraystretch}{1.1}
\begin{table}
\centering
{\begin{tabular}{K{3cm}K{2cm}K{2cm}K{2cm}}
 \hline
             &  LPJMM  &  LPCM       &  CSBM     \\
 \hline
 MaxPEAR          & 0.956 (5)   &  0.918 (5)  &  --       \\
 MinBinder        & 0.956 (5)   &  0.933 (6) &  --       \\
 GreedyEPL        & 0.938 (5)   &  0.918 (5)  &  --       \\
 Variational-EM   & --          & --          & 0.900 (5) \\
 \hline
\end{tabular}}
\label{table:rand-index-far}
\vspace{0.3cm}
\caption{ARI and numbers of estimated groups (in parentheses).}
\end{table}}

{\renewcommand{\arraystretch}{1.1}
\begin{table}
\centering
{\begin{tabular}{lccccc}
 \hline
          & true value  & LPJMM    & LPCM   & SNSM    & CSBM      \\
 \hline
 density       & 0.1008 & 0.1011   & 0.1008 & 0.0956  & 0.0965 \\
 transitivity  & 0.4822 & 0.4775   & 0.5453 & 0.2758  & 0.3563 \\
 assortativity & 0.8643 & 0.8510   & 0.8594 & 0.6540  & 0.7725 \\
 \hline
\end{tabular}}
\label{table:summary-statistics-far}
\vspace{0.3cm}
\caption{Means of the summary statistics.}
\end{table}}

\begin{figure}
\centering
\includegraphics[width=.8\linewidth]{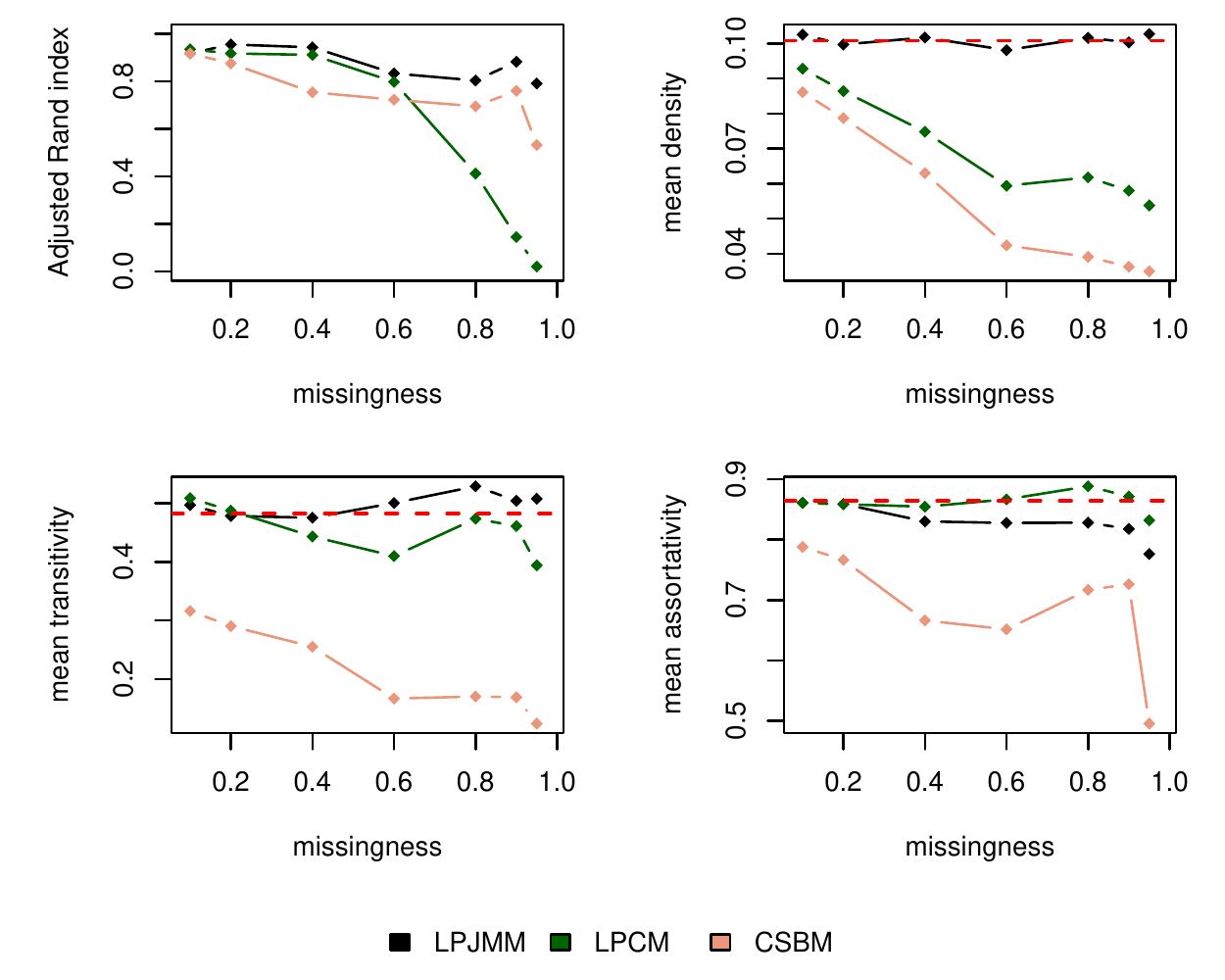}
\caption{The upper left plot shows ARI as the percentage of the missing edges increases from 10\% to 95\%.
In the rest of the three plots, the red horizontal dashed lines indicate the statistics summarized from the fully observed network. }
\label{fig:missing_scatter}
\end{figure}

As the level of missing edges rises from 10\% to 95\%, the advantage of LPJMM becomes more pronounced. To compare the models' ability to recover group membership, we use MaxPEAR to estimate $\mathbf{g}$ for LPJMM and LPCM. (The other two methods, MinBinder and GreedyEPL, will give a similar result.)
We also compare the summary statistics under different levels of missingness. (Note that SNSM is excluded from the comparison since it does not perform well even when data is complete.)
The results are shown in \cref{fig:missing_scatter}.

The upper left plot in \cref{fig:missing_scatter} reveals that LPCM closely matches LPJMM's performance when the missing data rate is below 60\%. However, as the amount of missing data increases, LPCM's performance deteriorates rapidly. Additionally, the density estimation from LPCM deviates significantly from the true density as the missing data proportion rises.
While CSBM's performance in terms of ARI appears less affected by the missing data rate, it struggles to recover all three summary statistics effectively.
Consequently, when dealing with missing data, LPJMM emerges as a considerably more robust approach.

\subsection{A network showing indiscernible clustering pattern}
\label{subsec-sim-3}

In this simulation scenario, we explore networks with well-blended group members. These networks do not show distinct clustering patterns in terms of the given group membership, which makes such membership more nominal than structural.

\begin{figure}
\centering
\includegraphics[width=.6\linewidth]{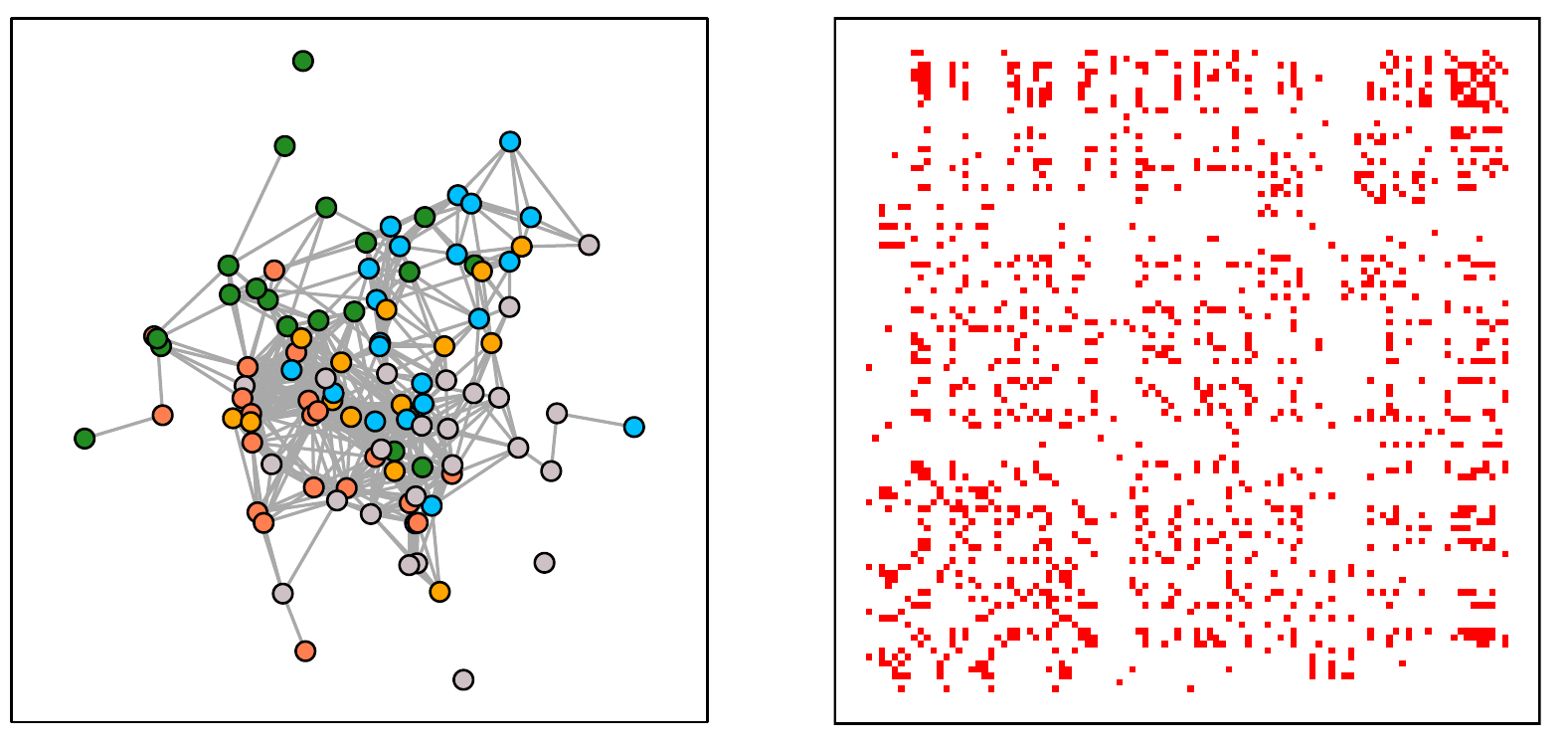}
\caption{Network visualization and heatmap of the adjacency matrix.}
\label{fig:network-close}
\end{figure}

Consider a network shown in \cref{fig:network-close}.
This network does not display a distinct clustering pattern, as indicated in \cref{fig:network-close}, along with a low assortativity coefficient, which is less than $0.1$ (see \cref{table:summary-statistics-close}).

{\renewcommand{\arraystretch}{1.1}
\begin{table}
\centering
{\begin{tabular}{K{3cm}K{2cm}K{2cm}K{2cm}}
 \hline
                  &  LPJMM      &  LPCM       &  CSBM     \\
 \hline
 MaxPEAR          & 0.072 (3)   &  0.083 (8)  &  --       \\
 MinBinder        & 0.096 (25)  &  0.058 (25) &  --       \\
 GreedyEPL        & 0.000 (1)   &  0.067 (3)  &  --       \\
 Variational-EM   & --          & --          &  0.075 (6) \\
 \hline
\end{tabular}}
\label{table:rand-index-close}
\vspace{0.3cm}
\caption{ARI and numbers of estimated groups (in parentheses).}
\end{table}}

{\renewcommand{\arraystretch}{1.1}
\begin{table}
\centering
{\begin{tabular}{lccccc}
 \hline
          & true value  & LPJMM    & LPCM   & SNSM    & CSBM      \\
 \hline
 density       & 0.1242 & 0.1246   & 0.1241 & 0.1213  & 0.1226 \\
 transitivity  & 0.5181 & 0.5286   & 0.5476 & 0.3875  & 0.3931 \\
 assortativity & 0.0840 & 0.0787   & 0.0742 & 0.0715  & 0.0481 \\
 \hline
\end{tabular}}
\label{table:summary-statistics-close}
\vspace{0.3cm}
\caption{Means of the summary statistics.}
\end{table}}

The lack of clear clustering makes it challenging to discern group memberships, which are minimally evident in the network data (see \cref{table:rand-index-close}).
However, the goodness-of-fit test given in \cref{table:summary-statistics-close} suggests that these models are still useful in recovering the network structure. Again, the results are consistent with the previous goodness-of-fit tests, with LPJMM slightly outperforming the other three models.

\subsection{A two-layer network}\label{subsec-sim-4}
\begin{figure}
\centering
\includegraphics[width=.6\linewidth]{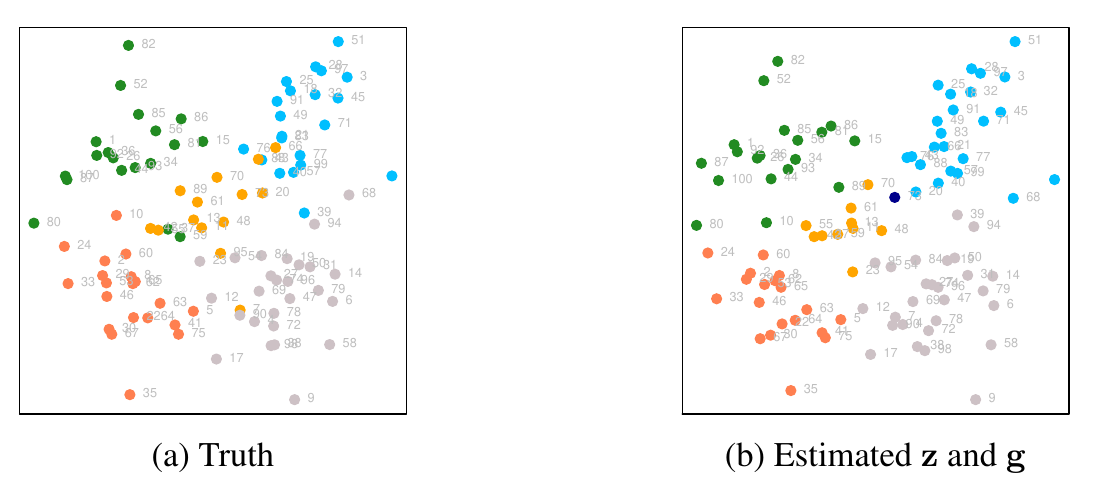}
\caption{(a): Points are plotted based on true latent position $\mathbf{z}$ and true group membership $\mathbf{g}$.
(b): Points are plotted using the estimated latent positions, and color represents the estimated group membership using the MaxPEAR method. }
\label{fig:sim2-z}
\end{figure}

Continue using the parameter setup in \cref{subsec-sim-1} and its generated network as the first layer,
we generate a second layer of the network with $a_2 = 3$, $b_2 = 1$, $\theta_2 = 4$.
As in \cref{subsec-sim-1}, we fit LPJMM with $K = 2$ and $H = 5$ and evaluate the model's ability to recover the group membership using ARI.
The results are given in
\cref{table:two-layer-rand-index}, which shows similar clustering estimates as in \cref{subsec-sim-1} where only one layer is considered.
The plot of the estimated latent position configurations is given in \cref{fig:sim2-z}~(b), which visualizes the model's recovery of latent positions and group membership.

{\renewcommand{\arraystretch}{1.1}
\begin{table}
\centering
{\begin{tabular}{K{1.5cm}K{2cm}K{2cm}K{2cm}}
\hline
                     &  MaxPEAR    & MinBinder   & GreedyEPL  \\
 \hline
             ARI     & 0.748 (6)   & 0.753 (12)  & 0.662 (4)  \\
 \hline
\end{tabular}}
\label{table:two-layer-rand-index}
\vspace{0.3cm}
\caption{ARI and numbers of estimated groups (in parentheses).}
\end{table}}

We also carry out the goodness-of-fit test and the result is given in
\cref{table:two-layer-summary-statistics},
which shows that LPJMM captures these structural features accurately, and the result for layer~1 is similar to the result in \cref{subsec-sim-1}.

{\renewcommand{\arraystretch}{1.1}
\begin{table}[h]
\centering
{\begin{tabular}{K{1.8cm}  K{1cm}K{1cm} | K{1cm}K{1cm} | K{1cm}K{1cm}}
\\
        & \multicolumn{2}{c|}{ \textbf{density} }
        & \multicolumn{2}{c|}{ \textbf{transitivity} }
        & \multicolumn{2}{c}{ \textbf{assortativity} }
\\
\cline{2-7}
        & truth   & mean    & truth  & mean    & truth  & mean  \\ 
\hline
layer 1 & 0.1531  & 0.1535  & 0.5049 & 0.5088  & 0.5512 & 0.5466 \\ 
layer 2 & 0.1024  & 0.1023  & 0.5477 & 0.5546  & 0.6923 & 0.6890 \\ 
\hline
\end{tabular}}
\label{table:two-layer-summary-statistics}
\vspace{0.3cm}
\caption{Means of the summary statistics.}
\end{table}}


Recall that $\theta_1$ and $\theta_2$ are of no direct interest since they are not identifiable. However, we are still interested in the ratio $\theta_1/\theta_2$ since it reflects the relative strength of borrowing information from the latent space of each layer.
Although $a_\ell$ and $b_\ell$ are of no direct interest, we pay attention to their signs, especially that of $b_\ell$ because different signs of $b_\ell$ have different interpretations of the effect of attributes as discussed in \cref{sec:model}.
We also assess the model’s ability to estimate parameters $\beta, \tau^2$, and $\sigma^2$ using posterior means and 95\% credible intervals.
The results are given in
\cref{table:sim-2-CI-table}.
Overall, the performance of LPJMM in recovering the true values of these model parameters is pretty well, except for $\tau^2$ and $\sigma^2$.
Both LPJMM and SNSM tend to underestimate $\sigma^2$ and overestimate $\tau^2$.
That is, the covariance of the attributes tends to be underestimated, and although $\tau^2$ is slightly overestimated, the variance of the attributes ($\tau^2 + \sigma^2$) still tends to be underestimated.

{\renewcommand{\arraystretch}{1.1}
\begin{table}
\centering
{\begin{tabular}{c K{2.6cm} K{2.6cm} K{4.2cm}}
\hline
                      & true value & posterior mean & 95\% credible interval   \\
 \hline
 $\theta_1/\theta_2$  & 0.680      & 0.653          & (0.579, \, 0.721)  \\
 $a_1$                & 5          & 5.01           & (4.719, \, 5.262) \\
 $a_2$                & 3          & 3.25           & (2.976, \, 3.572) \\
 $b_1$                & -2         & -1.919         & (-2.053, \, -1.766) \\
 $b_2$                & 1          & 1.058          & (0.901, \,  1.252) \\
 $\beta$              & 0          & -0.047         & (-1.027, \, 1.01 ) \\
 $\tau^2$             & 0.3        & 0.409          & (0.261, \, 0.592) \\
 $\sigma^2$           & 1          & 0.642          &  (0.230, \, 1.684)\\
 \hline
\end{tabular}}
\label{table:sim-2-CI-table}
\vspace{0.3cm}
\caption{Posterior means and 95\% credible intervals.}
\end{table}}

\subsection{A five-layer network}\label{subsec-sim-5}

\begin{figure}[h]
\centering
\includegraphics[width=.6\linewidth]{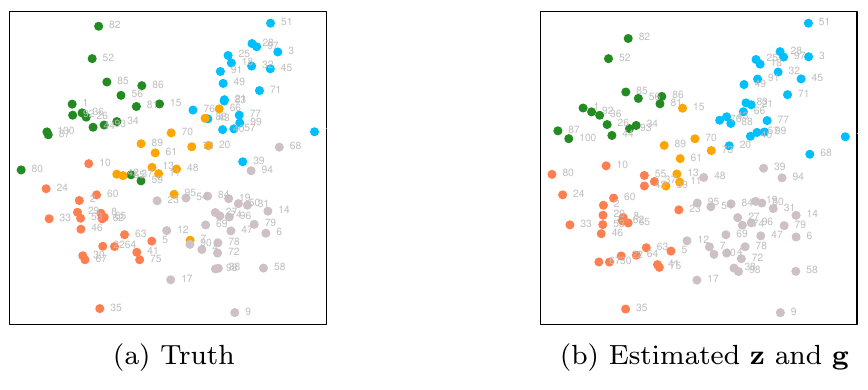}
\caption{(a): Points are plotted based on true latent position $\mathbf{z}$ and true group membership $\mathbf{g}$.
(b): Points are plotted using the estimated latent positions, and color represents the estimated group membership using the MaxPEAR method. }
\label{fig:sim5-z}
\end{figure}

In this section, we study a more complicated network by adding three more layers to the network in \ref{subsec-sim-4} to make it a five-layer network, and use the same model specification to assess the model fit of LPJMM. The results of the clustering estimates and the means of summary statistics are given in \cref{table:five-layer-rand-index,table:five-layer-summary-statistics} respectively.
The plot of the estimated latent position configurations is given in \cref{fig:sim5-z}~(b).

{\renewcommand{\arraystretch}{1.1}
\begin{table}[h]
\centering
{\begin{tabular}{K{1.5cm}K{2cm}K{2cm}K{2cm}}
\hline
                     &  MaxPEAR    & MinBinder   & GreedyEPL  \\
 \hline
             ARI     & 0.6928 (5)   & 0.7011 (11)  & 0.6414 (4)  \\
 \hline
\end{tabular}}
\label{table:five-layer-rand-index}
\vspace{0.3cm}
\caption{ARI and numbers of estimated groups (in parentheses).}
\end{table}}

{\renewcommand{\arraystretch}{1.1}
\begin{table}[h]
\centering
{\begin{tabular}{K{1.8cm}  K{1cm}K{1cm} | K{1cm}K{1cm} | K{1cm}K{1cm}}
\\
        & \multicolumn{2}{c|}{ \textbf{density} }
        & \multicolumn{2}{c|}{ \textbf{transitivity} }
        & \multicolumn{2}{c}{ \textbf{assortativity} }  \\
\cline{2-7}
        & truth   & mean    & truth  & mean    & truth  & mean  \\ 
\hline
layer 1 & 0.1531  & 0.1534  & 0.5049 & 0.5048  & 0.5512 & 0.5382 \\ 
layer 2 & 0.1024  & 0.1024  & 0.5477 & 0.5430  & 0.6923 & 0.6727 \\ 
layer 3 & 0.0648  & 0.0648  & 0.3661 & 0.4035  & 0.6977 & 0.7020 \\ 
layer 4 & 0.0543  & 0.0541  & 0.3892 & 0.3565  & 0.6744 & 0.7060 \\ 
layer 5 & 0.0169  & 0.0169  & 0.1421 & 0.1140  & 0.6930 & 0.7293 \\ 
\hline
\end{tabular}}
\label{table:five-layer-summary-statistics}
\vspace{0.3cm}
\caption{Means of the summary statistics.}
\end{table}}

Results of \cref{table:five-layer-rand-index}, assortativity estimates in \cref{table:five-layer-summary-statistics} and \cref{fig:sim5-z} indicate that as the network data becomes more complex with additional layers, the model's ability to identify clustering patterns slightly declines, with the ARI decreasing by approximately 0.05 compared to \cref{table:two-layer-rand-index} in the two-layer network setting.
However, the density and transitivity estimates in \cref{table:five-layer-summary-statistics} are similar to the results from \cref{subsec-sim-4}.

In this simulation, we are using the same Markov chain length as in previous simulations. However, it takes about 20~hours to generate the Markov chain, compared to about 8~hours for single-layer simulations.
A computational complexity analysis of the running time with respect to various layers, number of actors in the network and different model specifications (choice of $K$ and $H$) can be found in Appendix~\ref{app:running-time}.
The slight decrease in the model's performance in recovering the clustering pattern might due to the model's limitations in handling more complex network scenarios.
Alternatively, it could be due to the inefficiency and/or inadequacy of the Markov chain Monte Carlo algorithm in exploring the posterior distribution, which becomes more complex as the dimensions of the network data increase (e.g., with the addition of more layers).
To fully explore the model’s capabilities, including the goodness-of-fit and the recovery of clustering patterns, future studies could focus on developing a customized MCMC algorithm for the proposed model.

\section{Real data analysis}\label{sec:illustration}

In this section, we consider a three-layer network data set from a corporate law firm from $1988$-$1991$ in New England \cite{lazega}.
This network describes three types of relationships (namely, networks of advice, friendship, and coworker contacts) between 71 lawyers in the law firm.
Several actor attributes are also collected:
age,
gender,
seniority (years with the firm),
office (located in Boston, Hartford, or Providence),
practice (litigation or corporate law),
law school the lawyers attended (Harvard or Yale, University of Connecticut, or other universities) and
status (partner or associate).
A principal component analysis (PCA) is performed on age and seniority attributes, and the first principal component explains 89\% of the variance which is of no surprise since age and seniority are highly correlated with a correlation coefficient being $0.78$.
We chose the first principal component to be the attribute $\mathbf{x}$ and let $H=8$ since it is the largest number of clusters we expect in the network.
Then the model is fitted to the network using the same prior and Markov chain setup as in \cref{sec:simulation}.

\begin{figure}
\centering
\includegraphics[width=.75\linewidth]{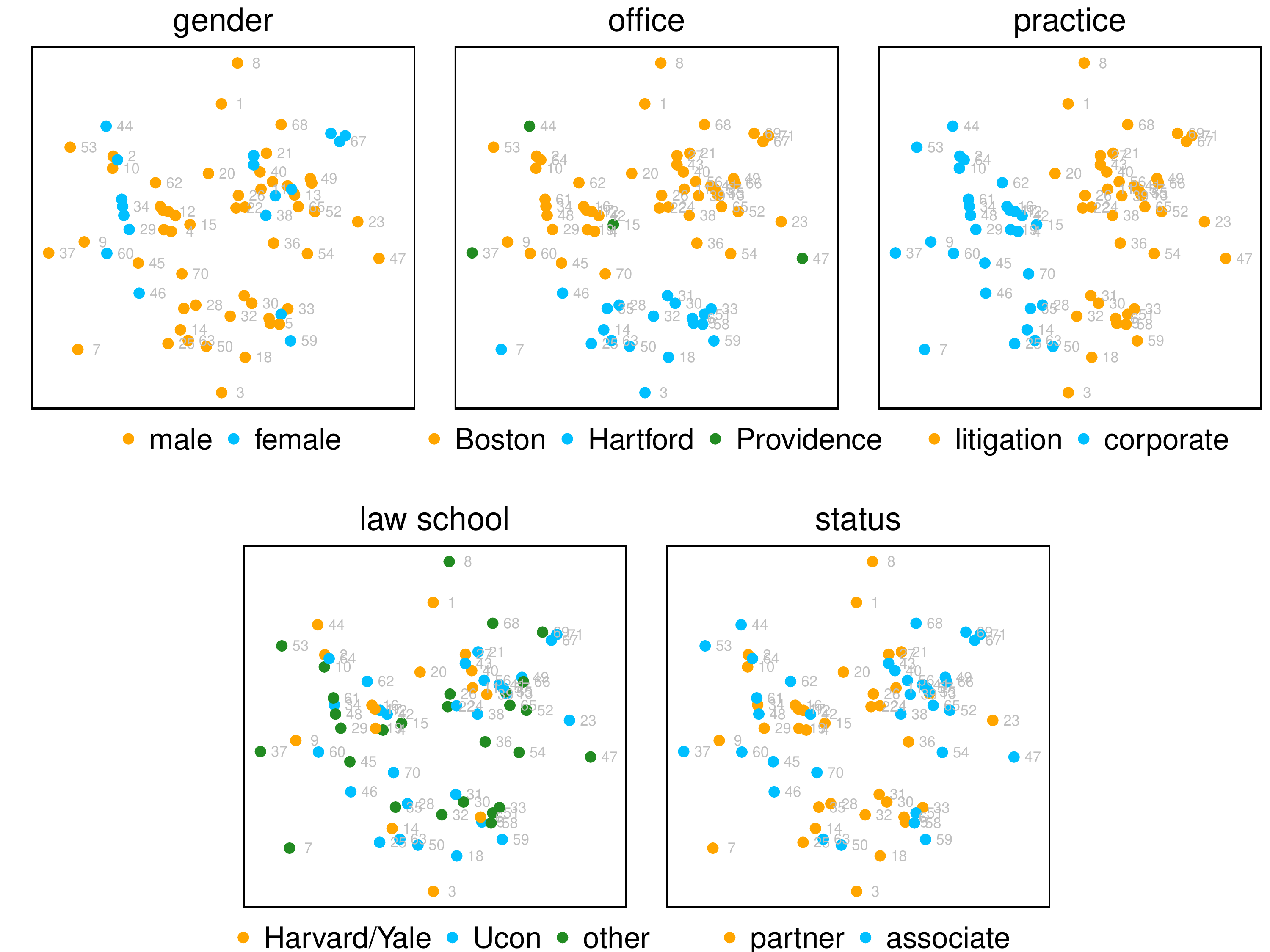}
\caption{Points in all plots are drawn based on the estimated latent positions $\mathbf{z}$ and are colored based on their categories in gender, office, practice, law school, and status.}
\label{fig:lazega-z-all-attributes}
\end{figure}

\begin{figure}
\centering
\includegraphics[width=.6\linewidth]{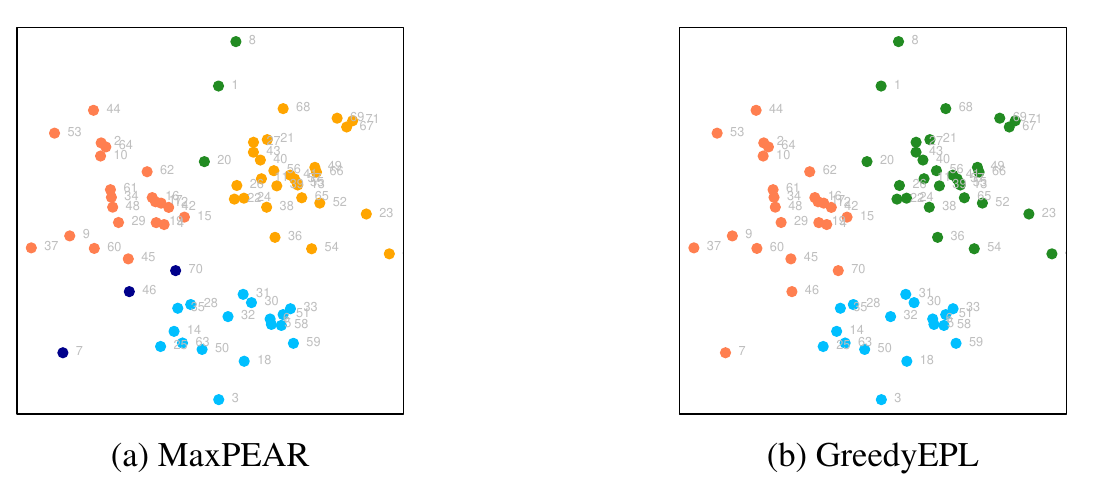}
\caption{Points are plotted using the estimated latent positions and color indicates the estimated group membership using MaxPEAR and GreedyEPL methods respectively. }
\label{fig:lazega-estmiated-g}
\end{figure}

The study of the Lazega network in this paper is meant to find out
how the three types of relations can be explained by the findings deduced from the model fitting.
We first visualize the estimated latent positions $\mathbf{z}$ colored by different categorical attributes (gender, office, practice, law school, and status) in \cref{fig:lazega-z-all-attributes}.
As we can see from these plots, the estimated positions $\mathbf{z}$ are well separated by the office (especially offices in Boston and Hartford) and practice.
Compare these plots with $\mathbf{z}$ colored by MaxPEAR and GreedyEPL estimated clustering $\mathbf{g}$ in \cref{fig:lazega-estmiated-g}, we can see that both estimated $\mathbf{g}$ roughly clusters lawyers into three groups: lawyers in Hartford office, litigation lawyers in Boston or Providence offices, and corporate lawyers in Boston or Providence offices.

\begin{figure}
\centering
\includegraphics[width=.7\linewidth]{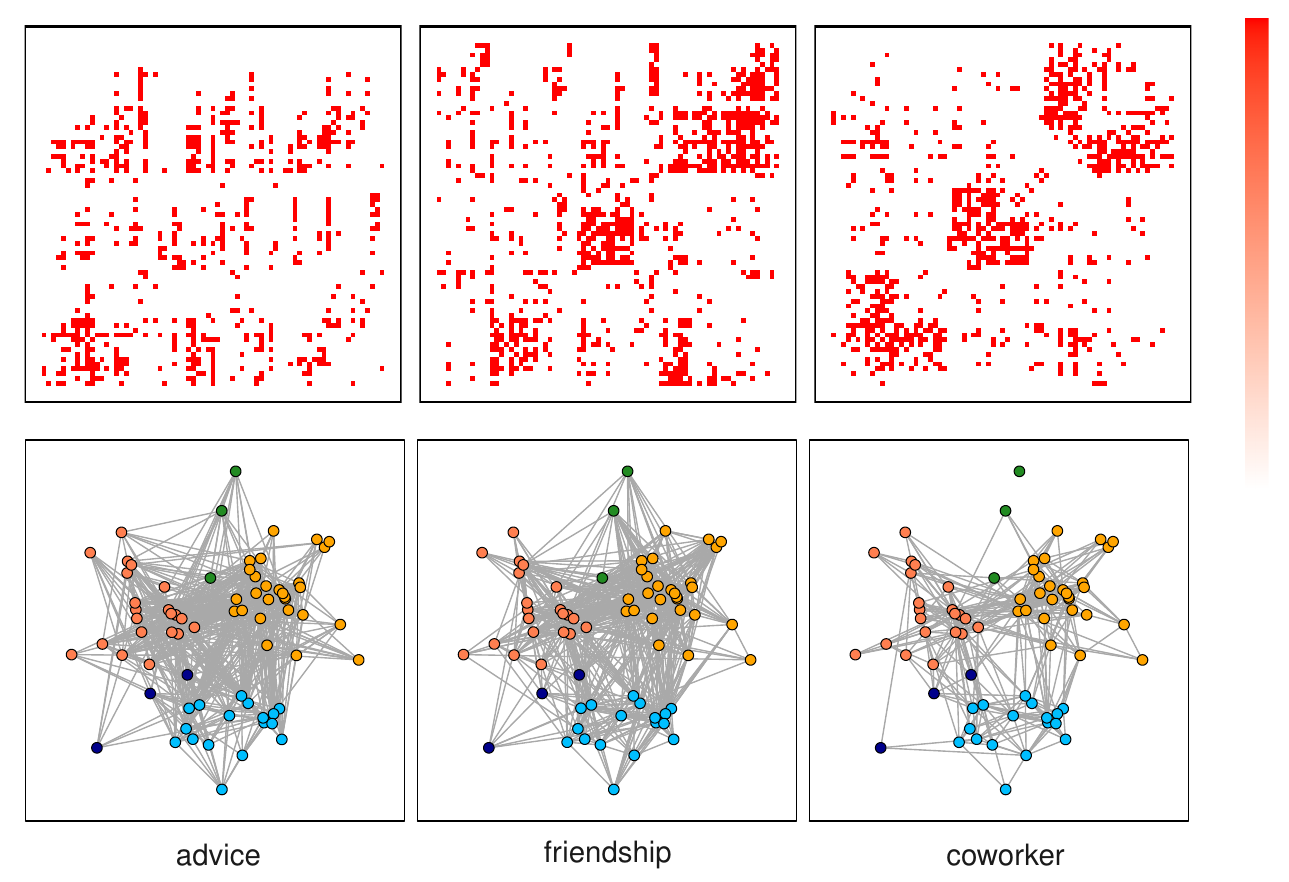}
\caption{Upper: Heatmaps of the adjacency matrices (where lawyers are reordered according to the MaxPEAR estimate of $\mathbf{g}$).
Lower: A visualization of the three layers based on the estimated $\mathbf{z}$ and color indicates the MaxPEAR estimate of $\mathbf{g}$.}
\label{fig:lazega-adj}
\end{figure}

Plots of adjacency matrices of the three layers (where lawyers are grouped by the MaxPEAR estimate of $\mathbf{g}$) and their corresponding networks are given in \cref{fig:lazega-adj}, where we could see that the coworker network shows the most estimated clustering pattern, while the advice network presents the least of such pattern, which could also be seen from the relative ratios of $\theta_\ell$s in \cref{table:lazega-CI-table}.
This means that lawyers from the same office and doing the same practice are more likely to become coworkers and friends, but who they seek advice from does not depend much on office locations and practices.
Furthermore, we can deduce from the posteriors of $b_\ell$ in \cref{table:lazega-CI-table} that these lawyers tend to seek advice from people of similar age (or seniority) since the posterior estimate of $b_1$ is negative. In contrast, lawyers of different ages (or seniority) are more likely to become friends and coworkers. This conclusion is in line with the assortativity coefficients found based on the nodal attributes (lawyer's age) given in \cref{table:assort-wrt-age}.

{\renewcommand{\arraystretch}{1.1}
\begin{table}
\centering
{\begin{tabular}{K{2cm}K{2cm}K{4cm}}
\hline
                      & posterior mean & 95\% credible interval   \\
 \hline
 $\theta_1/\theta_2$  & 0.3229         & (0.2352, \, 0.4152)  \\
 $\theta_1/\theta_3$  & 0.2035         & (0.1479, \, 0.2606)  \\
 $\theta_2/\theta_3$  & 0.6319         & (0.5536, \, 0.7198)  \\
 $b_1$                & -0.0986        & (-0.1401, \, -0.0579)  \\
 $b_2$                & 0.0708         & (0.0263, \, 0.1137)  \\
 $b_3$                & 0.133          & (0.0854, \, 0.186)  \\
 \hline
\end{tabular}}
\label{table:lazega-CI-table}
\vspace{0.3cm}
\caption{Posterior means and 95\% credible intervals.}
\end{table}}

{\renewcommand{\arraystretch}{1.1}
\begin{table}
\centering
{\begin{tabular}{K{2cm}K{1.5cm}K{1.5cm}K{1.5cm}}
\hline
               & advice   & friendship   & coworker   \\
 \hline
 assortativity & 0.2536   & -0.1107      & -0.1224 \\
 \hline
\end{tabular}}
\label{table:assort-wrt-age}
\vspace{0.3cm}
\caption{Assortativity coefficients based on lawyer's age.}
\end{table}}

\section{Discussion}\label{sec:discussion}

This paper presents a latent position model that extends LPCM and SNSM
to jointly model
network data and the nodal attributes and perform model-based clustering.
By jointly modeling the network and the attributes, we can describe how the attributes change over the network and explain how relations could be influenced by attributes.
LPJMM also provides an extension to multi-layer network settings on the assumption that all layers share the same latent position structure but with different strengths of borrowing such latent structure information.
We applied our method to four simulated networks and found LPJMM to give more satisfactory fits and is competitive in terms of goodness-of-fit and group detection compared with SNSM, LPCM, and CSBM.
In addition, the advantage of LPJMM is more pronounced when there is missing data in the network and LPJMM is shown to be much more robust than the other models.
LPJMM is also applied to a three-layer real network data set and we are able to draw reasonable conclusions from the modeling results.

We have suggested choosing the number of groups $H$ to be the largest number of groups that one is willing to accept in the network because we have found that varying the number of groups has almost no impact on the model fit and prediction outcome as long as it is in a reasonable range.
One could also fit the CSBM to the network first, and choose $H$ based on its estimated number of groups.
One problem we have not addressed in the paper is choosing the dimension of the latent space. This can be done by using Bayesian model selection like WAIC as in \cite{sosa:2022}.

Our model could be extended in several ways.
Firstly,
other extensions of our model to multi-layer settings could be considered.
For example, \citet{sosa:2022} assumed conditionally independent layer-specific latent positions,
whereas \citet{Multi-1:2022}
assumed that the latent position of an actor in all layers is $(d_0 + d_1)$-dimensional, where the first $d_0$ components of the latent position are the same across all layers, and only the last $d_1$ components are layer-specific.
Secondly, instead of assigning a user-specified number of groups $H$ to the model, we could learn the number of groups by using a Bayesian nonparametric approach with a Dirichlet Process prior to model community memberships
(see, e.g., \cite{lin:2021}).

LPJMM could also be extended to leverage multivariate covariates.
So far, we have limited ourselves to modeling univariate nodal attributes that are approximately Gaussian.
For continuous nodal attributes with more than one dimension, we have used the first principal component from the principal component analysis.
To take full advantage of high-dimensional nodal attributes, one could use multivariate spatial process modeling to replace \cref{model-level-2}.
Other extensions of more sophisticated spatial modeling include spatiotemporal modeling of attributes for time-varying networks, which would help to describe changes in actors over time.

\section*{Data availability statement}
The code and datasets generated and analyzed during the current study are available on GitHub:
\url{https://github.com/zjin3/LPJMM}.

\section*{Funding}
This work was supported by the National Science Foundation [grant number 2051911].



\begin{thebibliography}{58}
\providecommand{\natexlab}[1]{#1}
\providecommand{\url}[1]{\texttt{#1}}
\expandafter\ifx\csname urlstyle\endcsname\relax
  \providecommand{\doi}[1]{doi: #1}\else
  \providecommand{\doi}{doi: \begingroup \urlstyle{rm}\Url}\fi

\bibitem[Airoldi et~al.(2007)Airoldi, Blei, Fienberg, and Xing]{Airoldi:2007}
E.~M. Airoldi, D.~M. Blei, S.~E. Fienberg, and E.~P. Xing.
\newblock Mixed membership analysis of high-throughput interaction studies.
\newblock \emph{arXiv:0706.0294}, 2007.

\bibitem[Albert and Chib(1993)]{albert:chib:1993}
J.~H. Albert and S.~Chib.
\newblock { Bayesian analysis of binary and polychotomous response data}.
\newblock \emph{Journal of the American Statistical Association}, 88:\penalty0
  669--679, 1993.

\bibitem[Aldous(1985)]{aldous:1985}
D.~J. Aldous.
\newblock \emph{{Exchangeability and related topics}}.
\newblock Springer, Berlin, Heidelberg, 1985.

\bibitem[Amini et~al.(2019)Amini, Paez, and Lin]{lin:2021}
A.~A. Amini, M.~S. Paez, and L.~Lin.
\newblock Hierarchical stochastic block model for community detection in
  multiplex networks.
\newblock \emph{arXiv:1904.05330}, 2019.

\bibitem[Athreya et~al.(2017)Athreya, Fishkind, Tang, Priebe, Park, Vogelstein,
  Levin, Lyzinski, Qin, and Sussman]{athr:2017}
A.~Athreya, D.~E. Fishkind, M.~Tang, C.~E. Priebe, Y.~Park, J.~T. Vogelstein,
  K.~Levin, V.~Lyzinski, Y.~Qin, and D.~L. Sussman.
\newblock Statistical inference on random dot product graphs: a survey.
\newblock \emph{Journal of Machine Learning Research}, 18:\penalty0 1--92,
  2017.

\bibitem[Banerjee et~al.(2015)Banerjee, Carlin, and Gelf]{banerjee:2015}
S.~Banerjee, B.~Carlin, and A.~Gelf.
\newblock \emph{{Hierarchical Modeling and Analysis for Spatial Data}}.
\newblock CRC Press, 2nd edition, 2015.

\bibitem[Betancourt et~al.(2019)Betancourt, Rodríguez, and
  Boyd]{Brenda:finance:2019}
B.~Betancourt, A.~Rodríguez, and N.~Boyd.
\newblock {Modelling and Prediction of Financial Trading Networks: An
  Application to the New York Mercantile Exchange Natural Gas Futures Market}.
\newblock \emph{Journal of the Royal Statistical Society Series C: Applied
  Statistics}, 69\penalty0 (1):\penalty0 195--218, 2019.

\bibitem[Campedelli et~al.(2019)Campedelli, Cruickshank, and
  Carley]{Campedelli:2019}
G.~Campedelli, I.~Cruickshank, and K.~M. Carley.
\newblock {A complex networks approach to find latent clusters of terrorist
  groups}.
\newblock \emph{Applied Network Science}, 4\penalty0 (1):\penalty0 59, 2019.

\bibitem[Chiquet et~al.(2022)Chiquet, Donnet, and Barbillon]{pkg:sbm}
J.~Chiquet, S.~Donnet, and P.~Barbillon.
\newblock \emph{{$\mathtt{sbm}$: Stochastic Blockmodels}}, 2022.
\newblock URL \url{https://CRAN.R-project.org/package=sbm}.
\newblock R package version 0.4.4.

\bibitem[Ciminelli et~al.(2019)Ciminelli, Love, and Wu]{Ciminelli}
J.~T. Ciminelli, T.~Love, and T.~T. Wu.
\newblock {Social network spatial model}.
\newblock \emph{Spatial Statistics}, 29:\penalty0 129--144, 2019.

\bibitem[Diebolt and Robert(1994)]{diebolt:1994}
J.~Diebolt and C.~P. Robert.
\newblock {Estimation of finite mixture distributions through Bayesian
  sampling}.
\newblock \emph{Journal of the Royal Statistical Society. Series B},
  56:\penalty0 363--375, 1994.

\bibitem[Doreian et~al.(2004)Doreian, Batagelj, and Ferligoj]{Doreian:2004}
P.~Doreian, V.~Batagelj, and A.~Ferligoj.
\newblock \emph{{Generalized Blockmodeling (Structural Analysis in the Social
  Sciences}}.
\newblock Cambridge University Press, 2004.

\bibitem[Durante and Dunson(2018)]{dura:2018}
D.~Durante and D.~Dunson.
\newblock {Bayesian inference and testing of group differences in brain
  networks}.
\newblock \emph{Bayesian Analysis}, 13:\penalty0 29--58, 2018.

\bibitem[D’Angelo et~al.(2019)D’Angelo, Murphy, and Alf\`{o}]{dangelo:2019}
S.~D’Angelo, T.~B. Murphy, and M.~Alf\`{o}.
\newblock {Latent space modelling of multidimensional networks with application
  to the exchange of votes in Eurovision song contest}.
\newblock \emph{The Annals of Applied Statistics}, 13:\penalty0 900--930, 2019.

\bibitem[D’Angelo et~al.(2023)D’Angelo, Alf\`{o}, and Fop]{dangelo:2023}
S.~D’Angelo, M.~Alf\`{o}, and M.~Fop.
\newblock {Model-based clustering for multidimensional social networks}.
\newblock \emph{Journal of the Royal Statistical Society Series A: Statistics
  in Society}, 00:\penalty0 1--27, 2023.

\bibitem[Erd{\"o}s and R{\'e}nyi(1959)]{erdos:1959}
P.~Erd{\"o}s and A.~R{\'e}nyi.
\newblock {On random graphs. I.}
\newblock \emph{Publicationes Mathematicae (Debrecen)}, 6:\penalty0 290--297,
  1959.

\bibitem[Fosdick and Hoff(2015)]{fosd:2015}
K.~Fosdick and P.~D. Hoff.
\newblock {Testing and modeling dependencies between a network and nodal
  attributes,}.
\newblock \emph{Journal of the American Statistical Association}, 110:\penalty0
  1047--1056, 2015.

\bibitem[Frank and Strauss(1986)]{ERGM:1986}
O.~Frank and D.~Strauss.
\newblock {Markov graphs}.
\newblock \emph{Journal of the American Statistical Association}, 81:\penalty0
  832--842, 1986.

\bibitem[Fritsch(2022)]{pkg:mcclust}
A.~Fritsch.
\newblock \emph{{$\mathtt{mcclust}$: Process an MCMC Sample of Clusterings}},
  2022.
\newblock URL \url{https://CRAN.R-project.org/package=mcclust}.
\newblock R package version 1.0.1.

\bibitem[Fritsch and Ickstadt(2009)]{maxpear:2009}
A.~Fritsch and K.~Ickstadt.
\newblock {Improved criteria for clustering based on the posterior similarity
  matrix}.
\newblock \emph{Bayesian Analysis}, 4:\penalty0 367--391, 2009.

\bibitem[Gleditsch(2002)]{Gleditsch:2002}
K.~S. Gleditsch.
\newblock Expanded trade and gdp data.
\newblock \emph{The Journal of Conflict Resolution}, 46\penalty0 (5):\penalty0
  712--724, 2002.

\bibitem[Gollini and Murphy(2016)]{gollini:2016}
I.~Gollini and T.~B. Murphy.
\newblock {Joint modeling of multiple network views}.
\newblock \emph{Journal of Computational and Graphical Statistics},
  25:\penalty0 246--265, 2016.

\bibitem[Guha and Rodriguez(2021)]{guha:2021}
S.~Guha and A.~Rodriguez.
\newblock Bayesian regression with undirected network predictors with an
  application to brain connectome data.
\newblock \emph{Journal of the American Statistical Association}, 116\penalty0
  (534):\penalty0 581--593, 2021.

\bibitem[Handcock et~al.(2007)Handcock, Raftery, and Tantrum]{handcock:2007}
M.~S. Handcock, A.~E. Raftery, and J.~M. Tantrum.
\newblock {Model-based clustering for social networks}.
\newblock \emph{Journal of the Royal Statistical Society: Series A},
  170:\penalty0 301--354, 2007.

\bibitem[Hoff(2007)]{hoff:2007}
P.~Hoff.
\newblock Modeling homophily and stochastic equivalence in symmetric relational
  data.
\newblock In \emph{Advances in Neural Information Processing Systems}, pages
  657--664, 2007.

\bibitem[Hoff(2005)]{hoff:2005}
P.~D. Hoff.
\newblock {Bilinear mixed-effects models for dyadic data}.
\newblock \emph{Journal of the American Statistical Association}, 100:\penalty0
  286--295, 2005.

\bibitem[Hoff(2009)]{hoff:2009}
P.~D. Hoff.
\newblock {Multiplicative latent factor models for description and prediction
  of social networks}.
\newblock \emph{Computational and Mathematical Organization Theory},
  15:\penalty0 261--272, 2009.

\bibitem[Hoff et~al.(2002)Hoff, Raftery, and Handcock]{hoff:2002}
P.~D. Hoff, A.~E. Raftery, and M.~S. Handcock.
\newblock {Latent space approaches to social network analysis}.
\newblock \emph{Journal of the American Statistical Association}, 97:\penalty0
  1090--1098, 2002.

\bibitem[Hoover(1982)]{hoover:1982}
D.~N. Hoover.
\newblock Row-column exchangeability and a generalized model for probability.
\newblock \emph{Exchangeability in probability and statistics (Rome, 1981)},
  pages 281--291, 1982.

\bibitem[Kim and Leskovec(2012)]{kim:2012}
M.~Kim and J.~Leskovec.
\newblock {Multiplicative attribute graph model of real-world networks}.
\newblock \emph{Internet Mathematics}, 8:\penalty0 113--160, 2012.

\bibitem[Kolaczyk(2009)]{kolacyzk:2009}
E.~D. Kolaczyk.
\newblock \emph{{Statistical Analysis of Network Data}}.
\newblock Springer, 2009.

\bibitem[Kolaczyk and Cs{\'a}rdi(2020)]{kolaczyk:2020}
E.~D. Kolaczyk and G.~Cs{\'a}rdi.
\newblock \emph{{Statistical Analysis of Network Data with R}}.
\newblock Springer, 2nd edition, 2020.

\bibitem[Krivitsky and Handcock(2022)]{pkg:latentnet}
P.~N. Krivitsky and M.~S. Handcock.
\newblock \emph{{$\mathtt{latentnet}$: Latent Position and Cluster Models for
  Statistical Networks}}.
\newblock The Statnet Project (\url{https://statnet.org}), 2022.
\newblock URL \url{https://CRAN.R-project.org/package=latentnet}.
\newblock R package version 2.10.6.

\bibitem[Krivitsky et~al.(2009)Krivitsky, Handcock, Raftery, and
  Hoff]{kriv:2009}
P.~N. Krivitsky, M.~S. Handcock, A.~E. Raftery, and P.~D. Hoff.
\newblock {Representing degree distributions, clustering, and homophily in
  social networks with latent cluster random effects models}.
\newblock \emph{Social Networks}, 31:\penalty0 204--213, 2009.

\bibitem[Lau and Green(2007)]{minbinder:2007}
J.~W. Lau and P.~J. Green.
\newblock {Bayesian model based clustering procedures}.
\newblock \emph{Journal of Computational and Graphical Statistics},
  16:\penalty0 526–558, 2007.

\bibitem[Lazega(2001)]{lazega}
E.~Lazega.
\newblock \emph{{The Collegial Phenomenon: The Social Mechanisms of Cooperation
  Among Peers in a Corporate Law Partnership}}.
\newblock Oxford University Press, 2001.

\bibitem[Lee and Wilkinson(2019)]{Lee:2019}
C.~Lee and D.~J. Wilkinson.
\newblock {A review of stochastic block models and extensions for graph
  clustering}.
\newblock \emph{Applied Network Science}, 4:\penalty0 1--50, 2019.

\bibitem[Leger(2016)]{leger:2016}
J.-B. Leger.
\newblock {Blockmodels: A R-package for estimating in Latent Block Model and
  Stochastic Block Model, with various probability functions, with or without
  covariates}.
\newblock \emph{arXiv:1602.07587}, 2016.

\bibitem[Linkletter(2007)]{link:2007}
Crystal~D Linkletter.
\newblock \emph{Spatial process models for social network analysis}.
\newblock PhD thesis, Citeseer, 2007.

\bibitem[MacDonald et~al.(2020)MacDonald, Levina, and Zhu]{zhu:2020}
P.~W. MacDonald, E.~Levina, and J.~Zhu.
\newblock Latent space models for multiplex networks with shared structure.
\newblock \emph{arXiv:2012.14409}, 2020.

\bibitem[MacDonald et~al.(2022)MacDonald, Levina, and Zhu]{Multi-1:2022}
P.~W. MacDonald, E.~Levina, and J.~Zhu.
\newblock {Latent space models for multiplex networks with shared structure}.
\newblock \emph{Biometrika}, 109:\penalty0 683--706, 2022.

\bibitem[Matias and Robin(2014)]{matias:2014}
C.~Matias and S.~Robin.
\newblock {Modeling heterogeneity in random graphs through latent space models:
  a selective review}.
\newblock \emph{ESAIM: Proceedings and Surveys}, 47:\penalty0 55--74, 2014.

\bibitem[McGloin(2005)]{McGloin:2005}
J.~M. McGloin.
\newblock {Policy and Intervention Consideration of a Network Analysis of
  Street Gangs}.
\newblock \emph{Criminology \& Public Policy}, 4\penalty0 (3):\penalty0
  607--636, 2005.

\bibitem[Minhas et~al.(2019)Minhas, Hoff, and Ward]{minhas:2019}
S.~Minhas, P.~D. Hoff, and M.~D. Ward.
\newblock {Inferential approaches for network analysis: AMEN for latent factor
  models}.
\newblock \emph{Political Analysis}, 27:\penalty0 208--222, 2019.

\bibitem[Newman(2018)]{newman:2018}
M.~Newman.
\newblock \emph{{Networks}}.
\newblock Oxford, 2nd edition, 2018.

\bibitem[Nowicki and Snijders(2001)]{nowi:2001}
K.~Nowicki and T.~A.~B. Snijders.
\newblock {Estimation and prediction for stochastic block structures}.
\newblock \emph{Journal of the American Statistical Association}, 96:\penalty0
  1077–1087, 2001.

\bibitem[Plummer(2022)]{rjags:2022}
M.~Plummer.
\newblock \emph{{$\mathtt{rjags}$: Bayesian Graphical Models using MCMC}},
  2022.
\newblock URL \url{https://CRAN.R-project.org/package=rjags}.
\newblock R package version 4-13.

\bibitem[{R Core Team}(2021)]{R:2021}
{R Core Team}.
\newblock \emph{R: A Language and Environment for Statistical Computing}.
\newblock R Foundation for Statistical Computing, Vienna, Austria, 2021.
\newblock URL \url{https://www.R-project.org/}.

\bibitem[Rastelli(2021)]{pkg:GreedyEPL}
R.~Rastelli.
\newblock \emph{{$\mathtt{GreedyEPL}$: Greedy Expected Posterior Loss}}, 2021.
\newblock URL \url{https://CRAN.R-project.org/package=GreedyEPL}.
\newblock R package version 1.2.

\bibitem[Rastelli and Friel(2018)]{GreedyEPL:2018}
R.~Rastelli and N.~Friel.
\newblock {Optimal Bayesian estimators for latent variable cluster models}.
\newblock \emph{Statistics and Computing}, 28:\penalty0 1169–1186, 2018.

\bibitem[Salter-Townshend and McCormick(2017)]{town:2017}
M.~Salter-Townshend and T.~H. McCormick.
\newblock {Latent space models for multiview network data}.
\newblock \emph{The Annals of Applied Statistics}, 11:\penalty0 1217--1244,
  2017.

\bibitem[Schweinberger and Snijders(2003)]{schw:2003}
M.~Schweinberger and A.~B. Snijders.
\newblock {Settings in social networks: a measurement model}.
\newblock \emph{Sociological Methodology}, 33:\penalty0 307--341, 2003.

\bibitem[Sosa(2021)]{sosa:2021}
J.~Sosa.
\newblock {A review on latent space models for social networks}.
\newblock \emph{Revista Colombiana de Estadistica}, 44:\penalty0 171--200,
  2021.

\bibitem[Sosa and Betancourt(2022)]{sosa:2022}
J.~Sosa and B.~Betancourt.
\newblock A latent space model for multilayer network data.
\newblock \emph{Computational Statistics \& Data Analysis}, 169:\penalty0
  107432, 2022.

\bibitem[Volz and Meyers(2009)]{Volz:2009}
E.~Volz and L.~A. Meyers.
\newblock Epidemic thresholds in dynamic contact networks.
\newblock \emph{Journal of the Royal Society Interface}, 6\penalty0
  (32):\penalty0 233--241, 2009.

\bibitem[Wang et~al.(2019)Wang, Zhang, and Dunson]{wang:2019}
L.~Wang, Z.~Zhang, and D.~Dunson.
\newblock {Common and individual structure of brain networks}.
\newblock \emph{Annals of Applied Statistics}, 13:\penalty0 85--112, 2019.

\bibitem[Wang et~al.(2003)Wang, Chakrabarti, Wang, and Faloutsos]{wang:2003}
Y.~Wang, D.~Chakrabarti, C.~Wang, and C.~Faloutsos.
\newblock Epidemic spreading in real networks: an eigenvalue viewpoint.
\newblock In \emph{Proceedings of the 22nd International Symposium on Reliable
  Distributed Systems (SRDS ’03)}, pages 25--34, 2003.
\newblock \doi{10.1109/RELDIS.2003.1238052}.

\bibitem[Wang and Wong(1987)]{wang:1987}
Y.~J. Wang and G.~Y. Wong.
\newblock {Stochastic blockmodels for directed graphs}.
\newblock \emph{Journal of the American Statistical Association}, 82:\penalty0
  8--19, 1987.

\end{thebibliography}

\appendix

\section{Model Specifications for SNSM, LPCM and CSBM}\label{app:other_models}

Note that the original SNSM in \cite{Ciminelli} uses the logit link. In order to make a fair comparison, we also use the probit link in SNSM as in LPJMM.
The model specification for SNSM used in this paper is given as follows:
\begin{align*}
y_{i,j} \mid \mathbf{z}, \mathbf{x}, a_\ell, b_\ell, \theta_\ell
& \stackrel{\text{ind}}{\sim}
\mathrm{Ber} \big( \Phi(a + b |x_i - x_j| - \|z_i - z_j\|) \big)  \, , \\
\mathbf{x} \mid \mathbf{z}, \beta, \sigma, \tau, \phi
& \sim
\mathrm{N}_N (\beta\pmb{1}_N, \, \sigma^2 M (\mathbf{z},\phi) + \tau^2 I_N) \, ,
\end{align*}
and the priors are set to be the same as the priors in LPJMM (if possible). To be specific,
\begin{align*}
z_i & \stackrel{\text{i.i.d.}}{\sim} \mathrm{N}_2 (\pmb{0}, \, I_2)\, ,
\beta    \sim \mathrm{N} (0, 10^4)\, ,
\sigma^2 \sim \mathrm{InvG} (2, 1)\, ,
\tau^2   \sim \mathrm{InvG} (2, 1)\, ,
\phi     \sim \mathrm{U} (0,1)\, ,
\end{align*}
and the priors on the parameters in the probit regression tier are given by:
\begin{align*}
a  \stackrel{\text{i.i.d.}}{\sim} \mathrm{N} (0, 9)\, ,
\quad
b  \stackrel{\text{i.i.d.}}{\sim} \mathrm{N} (0, 9)\, .
\end{align*}
SNSM in this paper is implemented using JAGS.

The model specification for LPCM (see \cite{handcock:2007}) is given as the follows,
\begin{align*}
y_{i,j} \mid \mathbf{z}, \mathbf{x}, \beta_0, \beta_1
& \stackrel{\text{ind}}{\sim}
\mathrm{Ber} \big(\mathrm{logit}(\beta_0^\intercal x_{i,j} - \beta_1\|z_i - z_j\|) \big)  \, , \\
z_i \mid \pmb{\omega}, \pmb{\mu}, \pmb{\kappa}
& \stackrel{\text{i.i.d.}}{\sim}
\sum_{h=1}^5 \omega_h  \mathrm{N}_5 (\mu_h, \, \kappa_h^2 I_K)\, ,
\end{align*}
and we use the default priors given in the $\mathtt{latentnet}$ package for prior specifications.

We first introduce several notations before presenting CSBM \cite{leger:2016}.
Suppose there are $Q$ groups in the network.
Denote the $N \times Q$ group membership matrix as $\pmb{Z} = \{Z_{iq}\}$, and $Z_{iq} = 1$ if actor $i$ belongs to group $q$, $Z_{iq} = 0$ if otherwise. It is assumed that an actor can only belong to one group.
The model specification for CSBM is given as follows,
\begin{align*}
y_{i,j}
\mid
Z_i, Z_j, \mathbf{x}, \beta
& \stackrel{\text{ind}}{\sim}
\mathrm{Ber} \big(\mathrm{logit}(m_{q_i, q_j} + \beta^\intercal x_{i,j}) \big)  \, ,
\end{align*}
where
$Z_i$ is the $i$-th row of $\pmb{Z}$,
$q_i$ is the group membership for actor $i$
and the group effect $m_{q_i, q_j} \in \mathbb{R}$.

\section{Comparing Model Performances for Different Number of Groups}\label{app:compare-H}

\begin{figure}
\centering
\includegraphics[width=.7\linewidth]{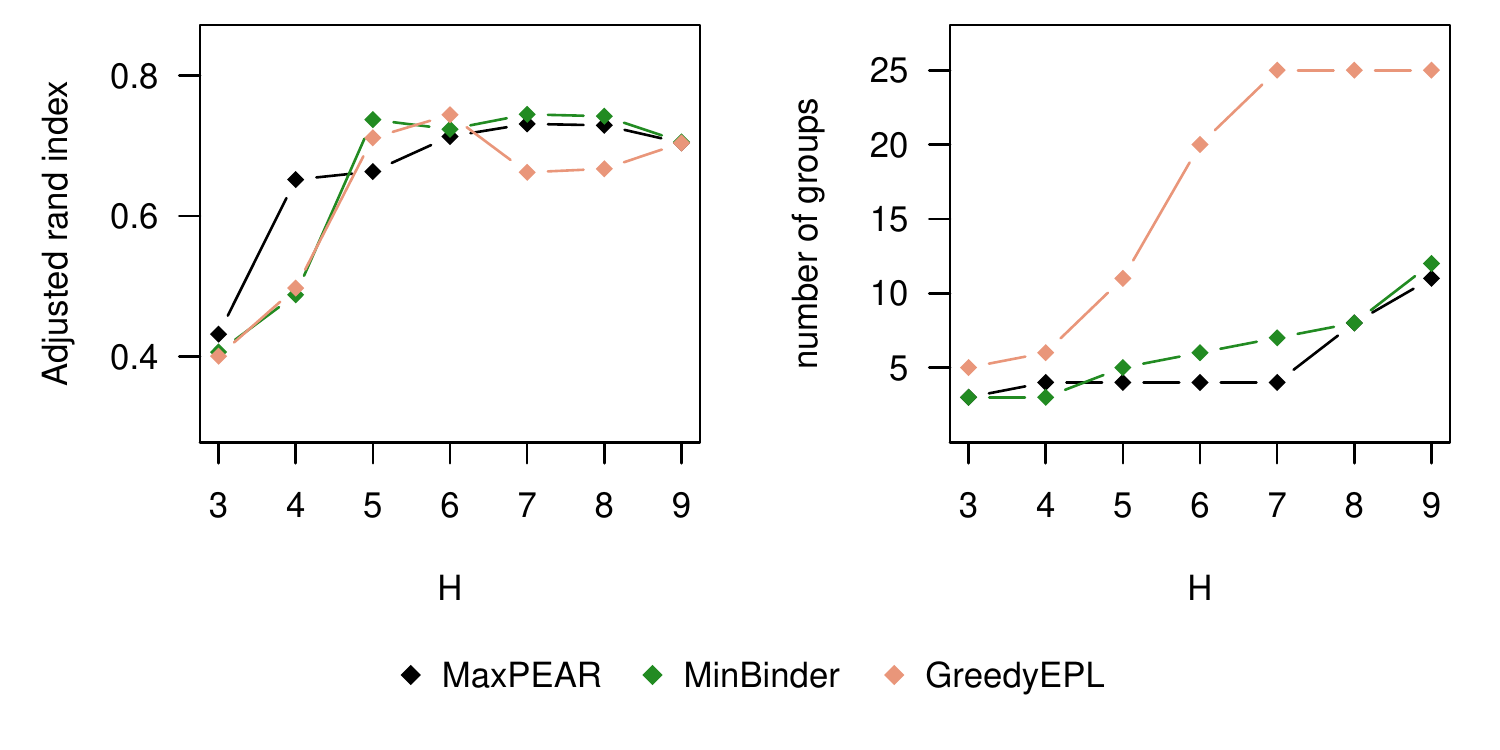}
\caption{Left: ARI of the clustering estimates found by using the MaxPear, MinBinder, and GreedyEPL methods.
Right: Estimated number of groups using the three methods.}
\label{fig:rand-n-multiple-H}
\end{figure}

We conduct a comparison of LPJMM with different $H \in \{3, 4, \dots, 9\}$ using the data set in \cref{subsec-sim-1}.
\cref{table:rand-multiple-H} presents ARI, and the results are similar for models that assume $H$ to be equal to or larger than the true number of groups (which is $5$ in this example). However, ARI for all three estimates is significantly smaller when the model assumes $H$ to be smaller than $5$.
Also, notice that the estimated number of groups increases with $H$.
Visualizations of how ARI and the estimated number of groups change over $H$ are given in \cref{fig:rand-n-multiple-H}.

{\renewcommand{\arraystretch}{1.1}
\begin{table}
\centering
{\begin{tabular}{lccc}
\hline
 $H$  & MaxPEAR      & MinBinder   & GreedyEPL   \\
\hline
 3    & 0.4067 (3)   & 0.4008 (5)   & 0.4321 (3)  \\
 4    & 0.4882 (3)   & 0.4977 (6 )  & 0.6521 (4)  \\
 5    & 0.7374 (5)   & 0.7115 (11)  & 0.6635 (4)  \\
 6    & 0.7237 (6)   & 0.7442 (20)  & 0.7134 (4)  \\
 7    & 0.7449 (7)   & 0.6624 (25)  & 0.7313 (4)    \\
 8    & 0.7422 (8)   & 0.6674 (25)  & 0.7293 (8)   \\
 9    & 0.7056 (12)  & 0.7041 (25)  & 0.7043 (11)  \\
\hline
\end{tabular}}
\label{table:rand-multiple-H}
\vspace{0.3cm}
\caption{ARI and the numbers of estimated groups (in parentheses).}
\end{table}}

\begin{figure}
\centering
\includegraphics[width=.95\linewidth]{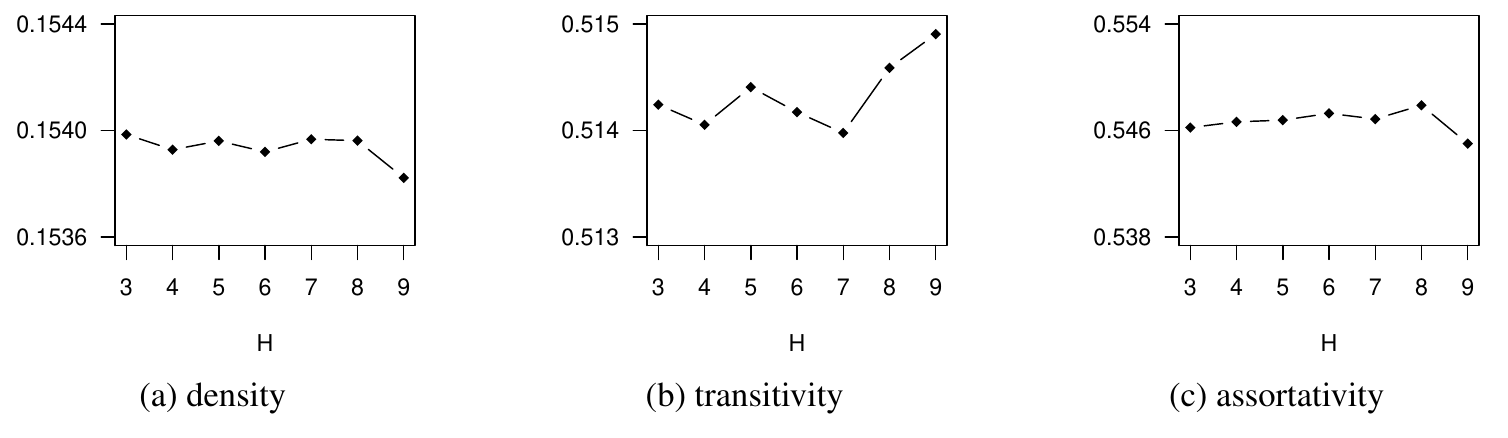}
\caption{The means of summary statistics for different $H$.}
\label{fig:GoF-multiple-H}
\end{figure}

The goodness-of-fit test outlined in \cref{sec:simulation} is also carried out here to compare the means of several summary statistics, which are plotted in \cref{fig:GoF-multiple-H}.
As we can see from the plots, the model's fit is not affected by the choice of $H$ even for $H$ smaller than the actual number of clusters in the network.

\section{Traceplots of Log-likelihood}\label{app:traceplots}

\begin{figure}[h]
\centering
\includegraphics[width=.95\linewidth]{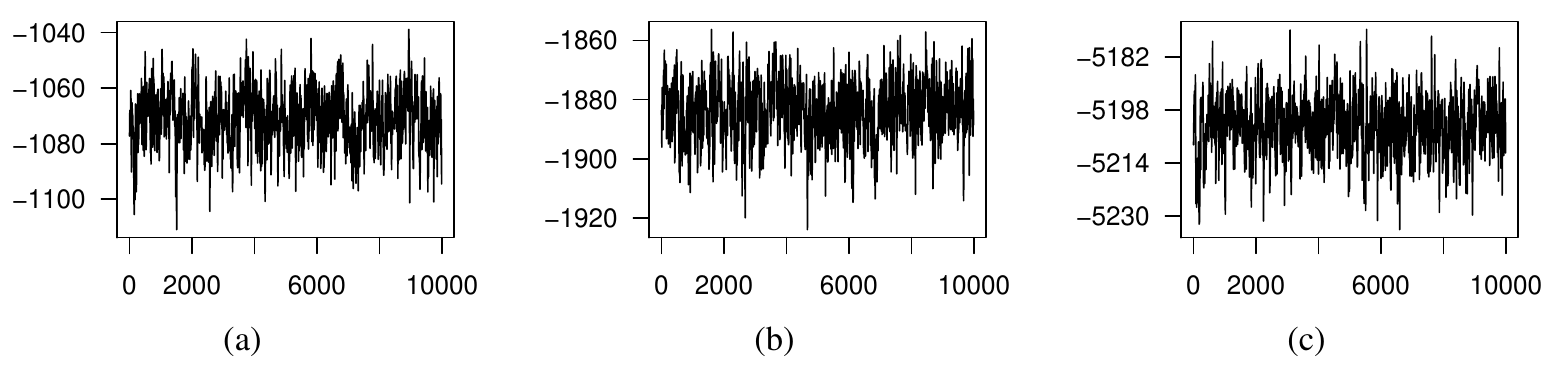}
\caption{From left to right: traceplots of the log-likelihood from \cref{subsec-sim-1,subsec-sim-4,sec:illustration} respectively.}
\label{fig:traceplot-loglikeli}
\end{figure}

The traceplots of the log-likelihood (after thinning the Markov chain every 10 iterations)
in simulation studies and real applications in \cref{sec:simulation,sec:illustration}
are given in \cref{fig:traceplot-loglikeli}.

\section{Visualizations of Results from LPJMM and LPCM }\label{app:compare-two-models}

\begin{figure}[h!]
\centering
\includegraphics[width=.95\linewidth]{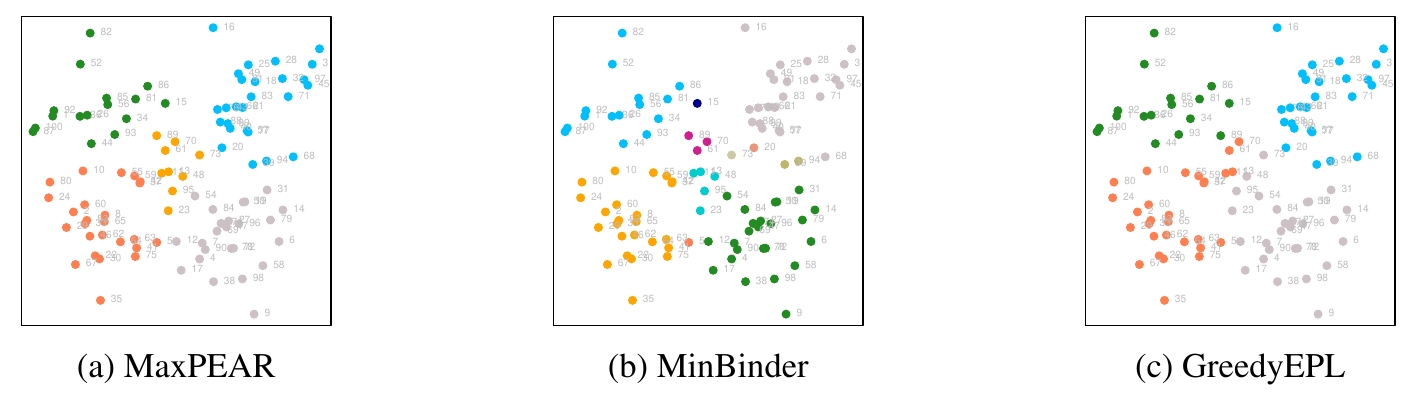}
\caption{Points are plotted based on the estimated latent position $\mathbf{z}$ and three estimated group memberships $\hat{\mathbf{g}}$ of LPJMM. }
\label{fig:Jin-z-all-g}
\end{figure}

\begin{figure}[h!]
\centering
\includegraphics[width=.95\linewidth]{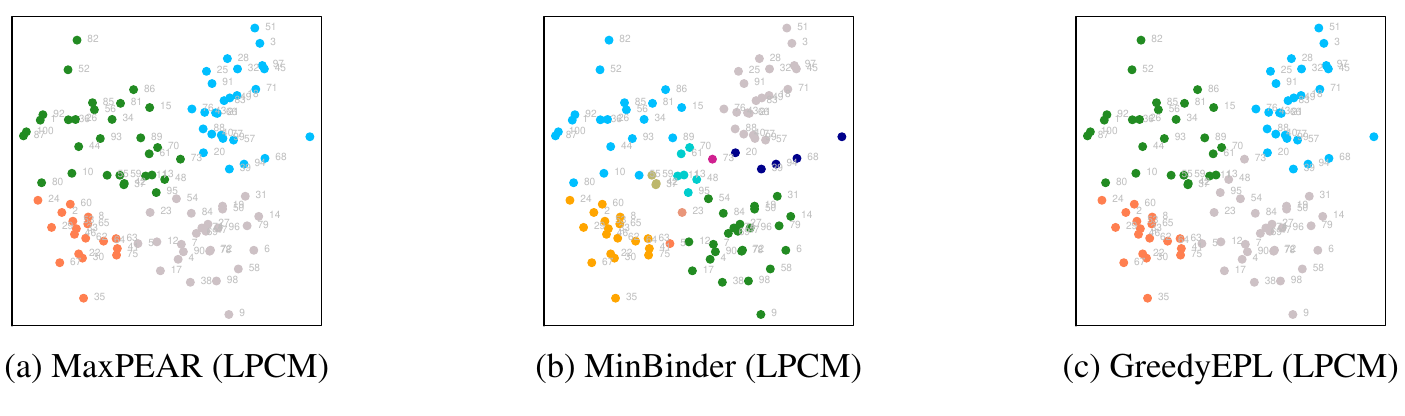}
\caption{Points are plotted based on estimated $\mathbf{z}$ and three estimated $\hat{\mathbf{g}}$ of LPCM.}
\label{fig:Hoff-all-g}
\end{figure}

Visualizations of the estimated latent positions and estimated group membership in \cref{subsec-sim-1} using the MaxPEAR, MinBinder, and GreedyEPL methods under
LPJMM
and LPCM are shown in \cref{fig:Jin-z-all-g,fig:Hoff-all-g} respectively.

\section{Computational Complexity Analysis}\label{app:running-time}
\begin{figure}[h]
\centering
\includegraphics[width=.99\linewidth]{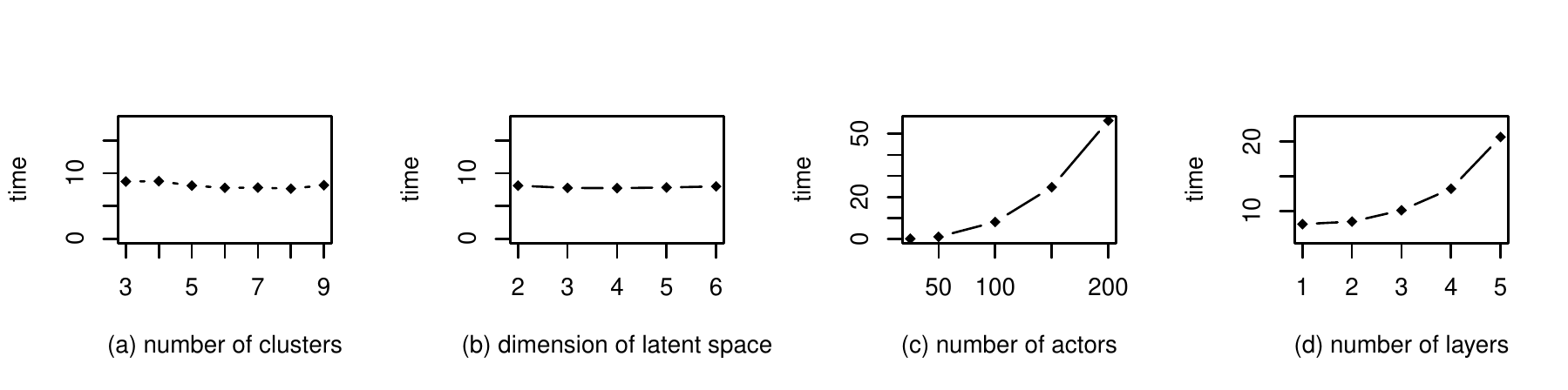}
\caption{ Running time (in hours) of the algorithm using the $\mathtt{rjags}$ package with 50000 iterations. (a) network is one-layer with 100 actors, and the dimension of latent space ($K$) is chosen to be 2. (b) network is one-layer with 100 actors and number of clusters ($H$) is chosen to be 5. (c) network is one-layer, and the model chooses $K=2$ and $H=5$. (d) network has 100 actors, and the model chooses $K=2$ and $H=5$.}
\label{fig:running_time}
\end{figure}

For all the Markov chains generated in this section, we utilize $20\,000$ iterations for adaption and $30\,000$ iterations for sampling the Markov chain.
The running time under different model specifications (different $K$ (i.e., dimension of latent space), different $H$ (i.e., number of clusters)), or when network data has different sizes (either different number of layers or actors) are shown in \cref{fig:running_time}.

From subplots (a) and (b) of \cref{fig:running_time}, it is evident that varying model specifications of $K$ and $H$ does not affect the running time. Subplots (c) and (d) illustrate an exponential trend in the running time as the number of actors or the number of layer increases.

\end{document}